\documentclass[12pt,a4wide]{article} 
 
\usepackage[]{amsmath}
\usepackage[]{graphicx}
\usepackage[]{latexsym}
\usepackage{geometry}
\def\beq{\begin{equation}} 
\def\eeq{\end{equation}} 
\def\barr{\begin{array}} 
\def\earr{\end{array}} 
\def\beqa{\begin{eqnarray*}} 
\def\eeqa{\end{eqnarray*}} 
\def\spa#1{\phantom{\fbox{\rule[-#1cm]{0cm}{0cm}}}} 
\font\mybb=msbm10 at 12pt 
\def\bb#1{\hbox{\mybb#1}}

\def\bE {\bb{E}} 
\def\bT {\bb{T}}

\def\b1 {\bb{1}} 

\def\lesssim{\mathrel{\hbox{\rlap{\hbox{\lower4pt\hbox{$\sim$}}}\hbox{$<$}}}} 
\def\gtrsim{\mathrel{\hbox{\rlap{\hbox{\lower4pt\hbox{$\sim$}}}\hbox{$>$}}}}

\renewcommand{\sp}{p\hspace{-.40em}/}
  
\begin{document} 

\vspace*{-.6in} \thispagestyle{empty}
\begin{flushright}
LPTENS--06/25\\
ROM2F/2006/16\\
\end{flushright}
\vspace{.2in} {\Large
\begin{center}
{\bf From Fundamental Strings to Small Black Holes}
\end{center}}
\vspace{.2in}
\begin{center}
Lorenzo Cornalba$^a$, Miguel S. Costa$^{b,c}$,
Jo\~ao Penedones$^{b,c}$ and Pedro Vieira$^{b,c,d}$
\\
\vspace{.3in} 
\emph{${}^a\,$Dipartimento di Fisica \& INFN,
Universit\'{a} di Roma ``Tor Vergata''\\
Via della Ricerca Scientifica 1, 00133, Roma, Italy}
\\
\vspace{.2in} 
\emph{${}^b\,$Departamento de F\'\i sica e Centro de F\'\i sica do Porto\\
Faculdade de Ci\^encias da Universidade do Porto\\
Rua do Campo Alegre 687,
4169--007 Porto, Portugal}
\\
\vspace{.2in} 
\emph{${}^c\,$Laboratoire de Physique Th\'eorique
de l'Ecole Normale Sup\'erieure\footnote{Unit\'e mixte du C.N.R.S.
et de l' Ecole Normale Sup\'erieure, UMR 8549.}\\
24 Rue Lhomond, 75231 Paris, France}
\\
\vspace{.2in} 
\emph{${}^d\,$Universit\'e Pierre et Marie Curie\\
4 Place Jussieu, 75005 Paris, France}

\end{center}

\vspace{.3in}

\begin{abstract}
We give evidence in favour of a string/black hole transition in the case of BPS fundamental string states of the
Heterotic string. Our analysis
goes beyond the counting of degrees of freedom and considers the evolution
of dynamical quantities in the process. As the coupling increases, the string states decrease their size 
up to the string scale when a small black hole is formed. We compute the
absorption cross section for several fields in both the black hole and the perturbative string phases. 
At zero frequency, these cross sections can be seen as order parameters
for the transition. In particular, for the scalars fixed at the horizon the cross section evolves 
to zero when the black hole is formed.
\end{abstract}

\newpage

\section{Introduction}

Since its original proposal \cite{Susskind}, the identification of fundamental strings with black holes has been widely 
recognized as a major challenge. The basic idea is to consider a Schwarzchild black hole of mass $M$ and entropy 
$S\sim GM^2$. If we decrease the gravitational coupling while keeping the entropy fixed, the hole radius
will decrease to the string scale. At this point,  the mass, which increased in the process, 
is related to the gravitational coupling by $G_*M_*\sim\sqrt{\alpha'}$.  
Furthermore, at this point, the entropy $S\sim \sqrt{\alpha'}M_*$, 
computed for a free string of mass  $M_*$, matches the black hole entropy. For smaller values of the coupling the 
black hole evolves into a weakly interacting string phase.

A major obstacle for a quantitative understanding of the string/black hole transition 
is  the uncontrollable self--energy corrections to the string mass in the process of increasing the coupling, i.e. 
of forming the hole \cite{Horowitz:1996nw,Damour}. BPS states have, on the other hand, a protected mass so that
the next logical step would be to use them to make precise quantitative statements about the transition.
However, supergravity black hole solutions corresponding to BPS strings are quite peculiar. 
These geometries have a horizon radius of the order of the string length, 
so that $\alpha'$ corrections to the supergravity effective action can not be neglected \cite{Dabholkar:1990yf}.
In the particular case of four--dimensional black holes of the Heterotic string, the corrected
near horizon geometry is $AdS_2\times S^2$ \cite{LopesCardoso:1998wt} and the corresponding Wald entropy \cite{Wald},
which includes the contribution of higher curvature terms, reproduces the microscopic string degeneracy \cite{Dabholkar}.
The main goal of this work is to go beyond this counting of degrees of freedom and to analyse the evolution
of dynamical quantities in the string/black hole transition. We shall study the absorption cross section for 
various fields, as well as the evolution of the size of string states.

\begin{figure}
\begin{picture}(0,0)(0,0)
\put(0,195){$\displaystyle{\frac{\langle{\cal R}^2\rangle}{\alpha'}}$}
\put(270,-4){$g^2N^{1/2}$}
\end{picture}
\centering\includegraphics[width=10cm]{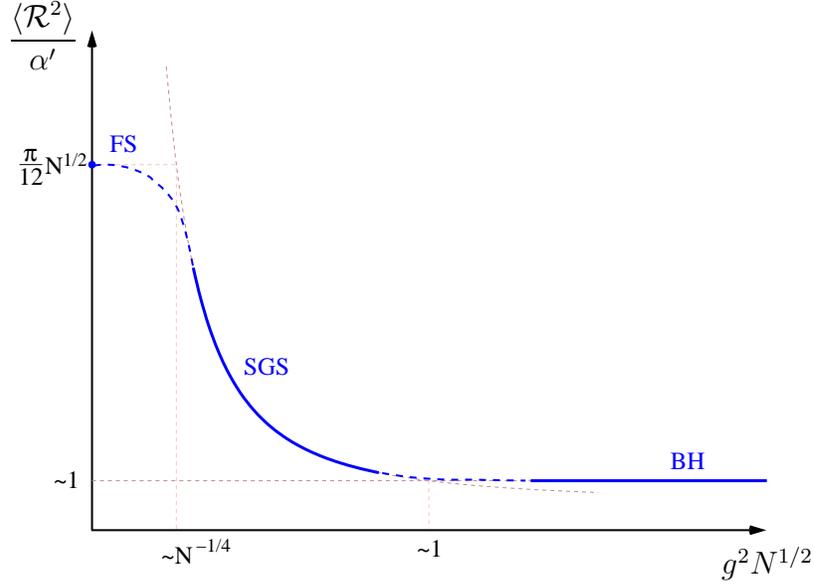} 
\caption{\small{Evolution of the string average size through the  transition.
The free string (FS) regime is appropriate for extremely small coupling $g^2\ll N^{-3/4}$. 
In the other extreme, for $g^2\gg N^{-1/2}$, the black hole (BH) phase prevails.
In between there is a self--gravitating string (SGS) phase.}}
\label{fig1}
\end{figure}

We consider a BPS state of the Heterotic string compactified on $S^1\times T^5$ 
carrying Ka\l u\.{z}a--Klein  and winding quantum numbers $(n,m)$ along the $S^1$ direction. 
In this controlled setup, let us briefly describe the string/black hole transition,
as shown in Figure \ref{fig1}. 
In the free string phase, for large values of the level $N=nm+1$, we will show that
the string average size is given by 
$$
\langle {\cal R}^2\rangle\simeq \frac{\pi}{12}\,\alpha'\sqrt{N} \ ,
$$
indicating that the usual random walk picture for very massive strings also holds
for BPS states. Turning on the string coupling $g$, the $\sqrt{N}$ bits of the string
will start to interact gravitationally, reducing the string size. 
We shall see that, above $g^2\sim N^{-3/4}$, the string is described by a self--gravitating 
string  phase \cite{Damour}, with 
$$
\langle {\cal R}^2\rangle \sim \frac{\alpha'}{g^4 N} \ .
$$
Finally, above $g^2\sim N^{-1/2}$, the large gravitational interaction collapses the string into 
a small black hole with fixed horizon radius (in the string frame) \cite{Dabholkar}
$$
{\cal R}_h^{\,2} \sim \alpha' \ ,
$$
which is independent of the charges $(n,m)$ and of the string coupling.

\begin{figure}
\begin{picture}(0,0)(0,0)
\put(-18,138){$\displaystyle{\frac{\sigma_{abs}}{G_N S}}$}
\put(270,-2){$g^2N^{1/2}$}
\end{picture}
\centering\includegraphics[width=10cm]{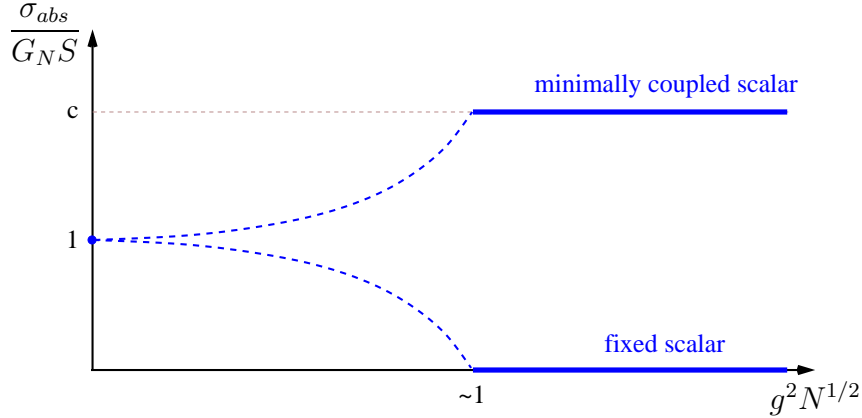} 
\caption{\small{Absorption cross section at $\omega=0$ as an order parameter.
Once  a black hole is formed the fixed scalar absorption cross section vanishes and
the minimally coupled scalar absorption cross section is proportional to the hole entropy 
times the Newton constant.
On the other hand, the perturbative string computation yields the same result for 
{\em all} massless particles. The precise behaviour of the cross section for all values of the coupling is not known
(as indicated by the dashed lines).}}
\label{fig2}
\end{figure}

Further evidence for the string/black hole transition  comes from the computation of the absorption cross
section. In the black hole phase we compute the low frequency behaviour of the absorption cross section
for a minimally coupled scalar and  for a fixed scalar\footnote{The fixed scalar result is
restricted to a particular background with constant $S^1$ radius.} obtaining, respectively,
$$
\sigma_{abs}\simeq c\, G_N S\ ,\ \ \ \ \ \ \ \ \sigma_{abs}\sim \omega^{\beta}\ ,
$$
where $c$ and $\beta>0$ are numerical constants of order $1$,
$G_N$ is the four--dimensional Newton constant and $S$ is the hole entropy. 
These results are valid to \textit{all} orders in $\alpha'$.
On the other hand, to leading order in perturbation theory
we determine the absorption cross section
for \textit{all} massless particles of the Heterotic string, finding the universal result
$$
\sigma_{abs}\simeq \frac{\pi}{2} g^2 \alpha'\sqrt{N} \simeq G_N S\  .
$$
These results support the physical picture explained above.
Moreover,
the vanishing of the  absorption cross section at $\omega=0$ for the fixed scalar
is directly related to the existence of a black hole and to the attractor mechanism.
Indeed, these scalars have their values at the horizon fixed by the hole charges
so that their oscillations, and therefore the infalling flux, are suppressed,
leading to the absence of absorption.
Since this is a generic behaviour provided
there is a horizon, we propose $\sigma_{abs}(\omega=0)$ as an
order parameter for the string/black hole (phase) transition (see Figure \ref{fig2}).

Finally, we shall see that both self--gravitating string and black hole phases can be
studied in the classical limit $g\rightarrow 0$ and $N\rightarrow \infty$ with $g^2\sqrt{N}$
fixed. Therefore, as explained in the Conclusion, one should be able to see the transition 
within the gravity framework.

In Section 2 we review the Heterotic BPS black hole geometry and compute the absorption cross section 
for both fixed and minimally coupled scalars in the gravity description.
Section 3 is devoted to the perturbative string CFT description. Firstly we compute
the absorption cross section for all massless fields. Then, at zero string coupling, we calculate the
average size of the string states corresponding to the black hole. We also estimate the string size for
small coupling. 
In Section 4 we give our conclusions. Finally, in Appendix A we study the low energy tunneling through a 
general one--dimensional Schr\"odinger potential, which is the problem one effectively needs to solve to study 
generic low energy scattering processes in spherically symmetric backgrounds.
In appendices B and C some longer calculations concerning the ensemble of 
string microstates are performed.

For earlier work on the relation between strings and BPS black holes see [8--11].
Recent work on the subject includes [12--29].

\section{Black hole phase}

\subsection{Review of the geometry}

We consider a black hole in Heterotic string theory compactified on 
$S^1\times T^5$. The only non--vanishing ten--dimensional fields are the
string metric $G_{ab}$, the dilaton field $\Phi $ and the NS 2--form gauge
potential $B_{ab}$. Neglecting $\alpha ^{\prime }$ corrections, the
ten--dimensional effective action for these fields is 
\[
\frac{1}{16\pi G_{N}^{^{\,(10)}}}\,\int d^{10}x\sqrt{-G}\,e^{-2\Phi }\left[
R+4(\partial \Phi )^{2}-\frac{1}{12}\,(dB)^{2}\right] \ , 
\]%
where the Newton constant is related to the string coupling and tension by 
\[
16\pi G_{N}^{^{\,(10)}}=(2\pi )^{7}g_{s}^{\,2}\,\alpha ^{\prime 4}\ . 
\]%
In our conventions the dilaton field vanishes at infinity, so that the
Newton constant has an explicit $g_{s}^{\,2}$ factor included.

The black hole will describe a fundamental string wrapping the $S^{1}$
circle, with Ka\l u\.{z}a--Klein momentum and winding numbers $(n,m)$. Therefore,
we consider a compactification to four dimensions with the ansatz 
\[
\begin{array}{c}
\displaystyle{ds_{10}^{\,2}=ds^{2}+e^{2\lambda }\left( dx^{4}+A_{\mu
}dx^{\mu }\right) ^{2}+e^{2\nu }\,ds^{2}\left( \bT^{5}\right) }\ ,
\phantom{\fbox{\rule[-0.5cm]{0cm}{0cm}}} \\ 
\displaystyle{2\Phi =2\phi +\lambda +5\nu \ ,\ \ \ \ \ \ \ \ \ \ \ \
B=C_{\mu }\,dx^{\mu }\wedge dx^{4}}\ ,\ \ \ 
\end{array}
\]
where greek indices run over the four non--compact space--time dimensions.
At infinity the radius of the $S^{1}$ circle is $R$ and the volume of the
five torus is $V_{5}$. We shall fix the asymptotic value of the scalar
fields $\lambda $ and $\nu $ to be zero. Then, the reduced effective action
has the form 
\[
\frac{1}{16\pi G_{N}}\,\int d^{4}x\sqrt{-G}\,e^{-2\phi }\left[ R+4(\partial
\phi )^{2}-(\partial \lambda )^{2}-5\left( \partial \nu \right) ^{2}-
\frac{e^{2\lambda }}{4}\,\left( dA\right) ^{2}-\frac{e^{-2\lambda }}{4}\,\left(
dC\right) ^{2}\right] \ . 
\]
The four--dimensional Newton constant is now given by
\[
8G_{N}=g^{2}\alpha ^{\prime }~, 
\]
where $g$ is the four--dimensional string coupling defined by 
\[
g^{2}=g_{s}^{\,2}\,\frac{(2\pi )^{6}\alpha ^{\prime 3}}{2\pi RV_{5}}\ .
\]
The field $\nu $ will only be important for us to study fluctuations of the
black hole geometry, so for now we set $\nu =0$.

The above action admits the BPS black hole solution \cite{Dabholkar:1990yf} 
\begin{equation}
\begin{array}{c}
\displaystyle{ds^{\,2}=-(FH)^{-1}\,dt^{2}+ds^{2}\left( \bE^{3}\right) }\ ,
\phantom{\fbox{\rule[-0.3cm]{0cm}{0cm}}} \\ 
\displaystyle{e^{2\phi }=(FH)^{-1/2}\ ,\ \ \ \ \ \ \ e^{\lambda }=\left( 
\frac{F}{H}\right) ^{1/2}\ ,}\phantom{\fbox{\rule[-0.5cm]{0cm}{0cm}}} \\ 
\displaystyle{A=F^{-1}\,dt\ ,\ \ \ \ \ \ \ C=H^{-1}\,dt}\ ,
\end{array}
\label{hole}
\end{equation}
where $F$ and $H$ are harmonic functions on $\bE^{3}$. For single--centered
black holes we have 
\[
F=1+\frac{\rho _{n}}{\rho }\ ,\ \ \ \ \ \ \ H=1+\frac{\rho _{m}}{\rho }\ , 
\]
where $\rho _{n}$ and $\rho _{m}$ are constants and $\rho $ is the radial
coordinate on $\bE^{3}$.

To quantize the black hole charges one can compute its ADM mass and equate
it to the mass of a string with momentum and winding $(n,m)$. Moving to the
Einstein metric $e^{-2\phi }G_{\mu \nu }$ we can compute the ADM mass 
\[
M=\frac{1}{4G_{N}}\,\left( \rho _{n}+\rho _{m}\right) =\frac{n}{R}+\frac{mR}{\alpha ^{\prime }}\ . 
\]
Thus, the constants $\rho _{n}$ and $\rho _{m}$ are given by 
\[
\rho _{n}=\frac{g^{2}\alpha ^{\prime }}{2}\,\frac{n}{R}\ ,\ \ \ \ \ \ \ \ \
\rho _{m}=\frac{g^{2}}{2}\,mR\ . 
\]
The classical limit, where the gravity description is adequate, consists in letting $g\rightarrow 0$
and $n,m\rightarrow\infty$ with the charges $\rho_n,\rho_m$ kept finite. As we shall see
below, {\it within this limit}, one needs to include curvature corrections to the 
effective gravity action.

Let us look at the region close to the black hole $\rho \ll \rho_n , \rho_m$. 
The string metric and remaining background fields become 
\begin{equation}
\begin{array}{c}
\displaystyle{ds^{2} = - \frac{\rho^2}{\rho_n\rho_m}\, dt^2 + d\rho^2 +
\rho^2 d\Omega^2\ ,}\phantom{\fbox{\rule[-0.4cm]{0cm}{0cm}}} \\ 
\displaystyle{e^{2\phi} = \frac{\rho}{\sqrt{\rho_n\rho_n}}\ , \ \ \ \ \ \ \
e^\lambda = \sqrt{\frac{\rho_n}{\rho_m}}}\ ,
\phantom{\fbox{\rule[-0.6cm]{0cm}{0cm}}} \\ 
\displaystyle{A = \frac{\rho}{\rho_n}\,dt \ , \ \ \ \ \ \ \ C = \frac{\rho}{%
\rho_m}\,dt}\ .%
\end{array}
\label{close}
\end{equation}
In the limit $\rho\rightarrow 0$ the geometry is singular, but the string
coupling goes to zero. For fixed angles on the 2--sphere the metric
is just the Rindler metric so that $\rho=0$ is a singular horizon. Since the
2--sphere has radius $\rho$, the Ricci scalar will diverge as $\rho^{-2}$
and string corrections will be important close to the horizon. Nevertheless,
when $\sqrt{\alpha^{\prime }}\ll\rho_n,\rho_m $, the above form of the
fields (\ref{close}) can be trusted in the region $\sqrt{\alpha^{\prime }}%
\ll \rho \ll \rho_n,\rho_m $.

To study the near--horizon geometry one needs to consider all $\alpha
^{\prime }$ corrections to the tree level supergravity action. It turns out
that, including a supersymmetric completion of curvature square terms in the
action, the near horizon geometry is given by $AdS_{2}\times S^{2}$ 
\cite{LopesCardoso:1998wt}. The corresponding Wald entropy formula \cite{Wald}
exactly reproduces the degeneracy of BPS string states with the same charges 
\cite{Dabholkar}.
We remark that in this scheme the field $\lambda$ diverges at the horizon,
but it is not the field associated to the radius of $S^1$ \cite{Sen}.
The same geometry can be obtained by
including only the Gauss--Bonnet correction in the Heterotic effective action 
\cite{SenGB}.
In this case, although the effective action is incomplete, the scalar field $\lambda$
is fixed at the horizon and the Wald entropy still reproduces the string degeneracy.
 There is now strong evidence that the $\alpha ^{\prime }$
corrected near--horizon solution is 
\begin{equation}
\begin{array}{c}
\displaystyle{ds^{2}=-\frac{2\rho ^{2}}{\rho _{n}\rho _{m}}\,dt^{2}+
\frac{\alpha ^{\prime }}{2\rho ^{2}}\,d\rho ^{2}+\frac{\alpha ^{\prime }}{2}
\,d\Omega ^{2}\ ,}\phantom{\fbox{\rule[-0.5cm]{0cm}{0cm}}} \\ 
\displaystyle{e^{2\phi }=\sqrt{\frac{\alpha ^{\prime }}{\rho _{n}\rho _{m}}}
\ ,\ \ \ \ \ \ \ e^{\lambda }=\sqrt{\frac{\rho _{n}}{\rho _{m}}}}\ ,
\phantom{\fbox{\rule[-0.6cm]{0cm}{0cm}}} \\ 
\displaystyle{A=\frac{2\rho }{\rho _{n}}\,dt\ ,\ \ \ \ \ \ \ 
C=\frac{2\rho }{\rho _{m}}\,dt\ .}
\end{array}
\label{nearhorizon}
\end{equation}%
The $AdS_{2}$ and $S^{2}$ radii squared are both $\alpha ^{\prime }/2$. At
the horizon, the square of the four--dimensional string coupling 
\[
g^{2}e^{2\phi }=\frac{2}{\sqrt{nm}}
\]
is completely fixed by the charges $(n,m)$. Thus, in the classical limit, loop corrections are
negligible throughout the whole space time. The other scalar field $\lambda 
$ is also fixed, with the radius of the $S^{1}$ circle at the horizon given
by 
\[
Re^{\lambda }=\sqrt{\alpha ^{\prime }\frac{n}{m}}\ .
\]%
Finally, the area of the horizon in the Einstein frame is 
\begin{equation*}
A_{h}=\pi \alpha ^{\prime }g^{2}\sqrt{nm} \simeq 2 G_N S \ .
\end{equation*}
Although the Wald entropy for the above near--horizon geometry exactly reproduces
the string degeneracy, we can not be completely certain of the precise numerical factors in the solution
because higher $\alpha'$--corections may change them. In the next section we shall assume this 
specific form of the metric in the cross section computations.

Sen \cite{Sen} gave strong evidence that an interpolating solution between
the near--horizon geometry (\ref{nearhorizon}) and the geometry close to the
black hole (\ref{close}) exists 
provided $\sqrt{\alpha^{\prime }}\ll\rho_n,\rho_m $. For this range of parameters one has 
\[
g^2 \sqrt{nm} \gg 1\ . 
\]
As described in the Introduction and in Section 3, increasing
the coupling to $g^2 \sqrt{nm} \sim 1$ the average size of the string states shrinks
to $\sqrt{\alpha^{\prime }}$. The string has then the same size of the above
horizon, therefore forming a black hole. This supergravity solution
describes the final stage of the transition.

For $\rho _{n},\rho _{m} \ll \sqrt{\alpha ^{\prime }}$, a solution
interpolating between the $AdS_{2}\times S^{2}$ near--horizon geometry and
the asymptotically flat geometry (\ref{hole}) is not known\footnote{Interpolating solutions between 
$AdS$ and infinity have been studied in \cite{Sen,Hubeny:2004ji}. In both papers these solutions present 
oscillations which were considered unphysical in \cite{Sen}. It was also argued that these 
oscillations could be removed by field redefinitions. Using the same setup we numerically
found interpolating solutions for all values of the charges. However, while for large charges,
$\sqrt{\alpha^{\prime }}\ll\rho_n,\rho_m $, the solution has small oscillations around the physical background,
for small charges, $\rho _{n},\rho _{m} \ll \sqrt{\alpha ^{\prime }}$, the oscillations dominate the
asymptotic behaviour of the field.}. This range of
parameters corresponds to the weakly coupled string phase 
\[
g^{2}\sqrt{nm}\ll 1\ , 
\]%
where the size of string states is much larger than the radius of the $AdS$
horizon. Therefore we believe that such an interpolating solution does not
exist. This is a very interesting open problem on which we shall comment in
the Conclusion.

\subsection{Absorption cross section}

We shall study the absorption cross section of massless fields by the
Heterotic black hole just described. In particular, we concentrate our
attention on minimally coupled scalars, like the fluctuation $\nu $ of the
volume of the internal $T^{5}$, and on fixed scalars, like the field $\lambda $ 
associated to the $S^{1}$ radius.

\subsubsection{Equations of motion}

Near infinity, where the curvature is small and $\alpha ^{\prime }$
corrections are negligible, the equations of motion for the field
fluctuations can be conveniently derived from the effective $4$--dimensional
action in the Einstein frame
\[
\frac{1}{16\pi G_{N}}\,\int d^{4}x\sqrt{-G}\,\left[ R-2(\partial \phi
)^{2}-(\partial \lambda )^{2}-5\left( \partial \nu \right) ^{2}-
\frac{e^{-2\phi +2\lambda }}{4}\,\left( dA\right) ^{2}-\frac{e^{-2\phi -2\lambda }
}{4}\,\left( dC\right) ^{2}\right] \ . 
\]
Thus, far from the black hole, the field $\nu $ is minimally coupled and
obeys the massless Klein--Gordon equation $\Box \nu =0$. In general,
however, the equations of motion change due to $\alpha ^{\prime }$
corrections, even though they  will still be invariant under
rescaling of the $T^{5}$ and $S^{1}$ volumes,  T--duality
transformations and shifts of the dilaton field. More precisely, the corrected effective action 
is invariant under the set of transformations
\begin{eqnarray}
&&\nu \rightarrow \nu +\alpha \ ;  \nonumber\\
&&\lambda \rightarrow \lambda +\beta \ ,\ \ \ \ \ \ \ dA\rightarrow
e^{-\beta }dA\ ,\ \ \ \ \ \ dC\rightarrow e^{\beta }dC\ ; \label{trans}\\
&&\nu \rightarrow -\nu \ ;  \nonumber \\
&&\lambda \rightarrow -\lambda ~,~\ \ \ \ \ \ \ \  dA\leftrightarrow dC\ ;\label{SoneTduality}\\
&&\phi \rightarrow \phi + \gamma \ , \ \ \ \ \ \ \ \ G_N \rightarrow e^{-2\gamma}G_N \ ; \label{dilshift}
\end{eqnarray}
with $\alpha ,\beta $ constants.
The cross section computations here performed rely on a field representation such that the effective action
has the above symmetries explicitly realized. We shall assume that the
geometry (\ref{nearhorizon}) is a solution of such an effective action.

Let us first focus on the field $\nu$. The effective action will depend
only on derivatives of the field and will be even in $\nu $. This fact
implies that fluctuations of $\nu $ do not mix with other fluctuations. Since the 
$AdS_{2}\times S^{2}$ geometry is a product of maximally
symmetric spaces, the near--horizon background fields respect the symmetries of this space.
Therefore, the only tensors available to define a differential equation near the horizon are
the metric and volume forms. Focusing on spherically symmetric
perturbations, the equation for the scalar field $\nu $ must be of the form 
\begin{equation}
f(\alpha ^{\prime }\Box )\,\nu =0\ ,  \label{correctedKG}
\end{equation}%
where the function $f(x)=f_{1}x+f_{2}x^{2}+\cdots $ depends, in principle,
on the hole charges. Notice that (\ref{correctedKG}) has no mass term due to
the symmetry $\nu \rightarrow \nu +\alpha$. Hence solutions of the
Klein--Gordon equation $\Box \nu =0$ are still solutions of the generic
equation (\ref{correctedKG}) and we expect those to be the physical
solutions. Possible other solutions to (\ref{correctedKG}) should be removed
after field redefinitions.
We conclude that the near--horizon action for s--wave physical fluctuations of the field $\nu$ 
has the generic form
\begin{equation}
\frac{5}{16\pi G_N} \int d^4 x \sqrt{-G}\,e^{-2\phi}\,K^{\alpha \beta} \partial_{\alpha} \nu \partial_{\beta} \nu \ ,
\label{nhaction}
\end{equation}
written in the string frame, and where $2 K_{\alpha \beta} = c\,G_{\alpha \beta} $.
We now argue that the proportionality constant $c$ is of order unity and independent of the hole charges.
In fact, rescaling the time coordinate  and  using the symmetries (\ref{trans}) and (\ref{dilshift})
one can remove the dependence of the near--horizon solution (\ref{nearhorizon}) on $\rho_n$ and $\rho_m$ 
\cite{Sen:1995in}.
In this new solution the constant $c$ must be independent of the hole charges.
This result holds in the original solution because both the metric and $K_{\alpha \beta}$  
are invariant under symmetries (\ref{trans}) and (\ref{dilshift}).

The study of fluctuations of the fixed scalars $\lambda $ and $\phi $ and of
the metric is more subtle since these fields are in general coupled to each
other. However, if the two charges $\rho _{n}$ and $\rho _{m}$ are equal,
the $S^{1}$ circle can have a constant background radius and the s--wave
fluctuations of the field $\lambda $ decouple from other fluctuations \cite%
{Kol}. Let us first recall that, in general, the equations of motion for the
gauge fields $A$, $C$ can always be put into the form 
$d{\mathcal F}=d{\mathcal H}=0$, where the two--forms ${\mathcal F}$,$~{\mathcal H}
$ are given, to leading order in $\alpha ^{\prime }$, by
\[
{\mathcal F}\simeq e^{2\lambda -2\phi }\star dA~,~\ \ \ \ \ \ \ \ \ \ \ 
{\mathcal H}\simeq e^{-2\lambda -2\phi }\star dC~,
\]%
and \ transform under (\ref{trans}) as ${\mathcal F}\rightarrow e^{\beta }%
{\mathcal F}$, ${\mathcal H}\rightarrow e^{-\beta }{\mathcal H}$.
For spherically symmetric solutions (and their fluctuations), the forms ${\mathcal F},~{\mathcal H}$ 
are constant along the transverse sphere
\[
{\mathcal F}=\rho _{n}~\epsilon \left( S^{2}\right) ~,~\ \ \ \ \ \ \ \ \ ~%
{\mathcal H}=\rho _{m}~\epsilon \left( S^{2}\right) ~,
\]
with $\epsilon \left( S^{2}\right)$ the volume form of  $S^{2}$.
Furthermore, their value is fixed by the overall charges $\rho _{n}$, $\rho _{m}$ of the
configuration. In particular, we will study fluctuations around a specific
extremal solution with
\[
\lambda =0,~\ \ \ \ \ \ \ \ \ \ \ \ \ \ \ A=C~,
\]
which is invariant under the T--duality transformation (\ref{SoneTduality}),
so that  $\rho _{n}=\rho _{m}$.
We will denote with $a$ the odd fluctuation of the gauge fields $\delta A=a$%
, $\delta C=-a$. The fields $\lambda $ and $a$ are the only fluctuations
which are odd under (\ref{SoneTduality}) and this implies that their linear
terms mix only among each other. Therefore, the relevant quadratic part of
the action reads
\[
\frac{1}{16\pi G_{N}}\int ~\left( \frac{1}{2}\,\lambda {\mathcal D}_{4}{\mathcal \lambda} 
+da\wedge {\mathcal D}_{2}\lambda -\frac{1}{2}\,da\wedge {\mathcal D}_{0}da\right) ~,
\]%
where the ${\mathcal D}_{i}$ are differential operators whose leading terms
in the $\alpha ^{\prime }$ expansions are given by 
\begin{eqnarray*}
&&{\mathcal D}_{4} \simeq 2d\star d-4e^{-2\phi }~dA\wedge \star dA~, \\
&&{\mathcal D}_{2} \simeq -4e^{-2\phi }\left( \star dA\right) \wedge ~, \\
&&{\mathcal D}_{0} \simeq 2e^{-2\phi }\star ~.
\end{eqnarray*}
The equation of motion for $a$ is given by $d\left( {\mathcal D}_{0}da-%
{\mathcal D}_{2}\lambda \right) =0$. As discussed above, for spherically
symmetric fluctuations this equation is solved by
\[
{\mathcal D}_{0}da-{\mathcal D}_{2}\lambda =\delta \epsilon \left(S^{2}\right) ~,
\]
where $\delta =\delta \rho _{n}-\delta \rho _{m}$ is the change in the
charge of the solution. We shall focus on perturbations which do not change
the charges so that $\delta =0$. Inverting the operator ${\mathcal D}_{0}$ in
power series in $\alpha ^{\prime }$ we may solve for $da$ as 
\[
da=\frac{1}{{\mathcal D}_{0}}\,{\mathcal D}_{2}\lambda ~.
\]
Defining the operator ${\mathcal D}_{2}^{\dagger }$ by $\int da\wedge 
{\mathcal D}_{2}\lambda =\int \lambda \wedge {\mathcal D}_{2}^{\dagger }\,da$\ ,
so that
\[
{\mathcal D}_{2}^{\dagger }\simeq -4e^{-2\phi }\left( \star dA\right) \wedge
~,
\]
the equation of motion for $\lambda $ reads ${\mathcal D}_{4}{\mathcal \lambda
+D}_{2}^{\dagger }\,da=0$ and can be recasted in the final form 
\[
{\mathcal D}\lambda =0~,
\]
with 
$\displaystyle{{\mathcal D}={\mathcal D}_{4}{\mathcal +D}_{2}^{\dagger }\,
\frac{1}{{\mathcal D}_{0}}\,{\mathcal D}_{2}}$\ .
To leading order in $\alpha ^{\prime }$ the fluctuation equation for $\lambda $ reads
\begin{equation}
\Box \lambda \simeq 2\rho _{n}^{2}~\frac{e^{2\phi }}{r^{4}}~\lambda ~,
\label{uncorrected}
\end{equation}
where $r$ is the radius of the $S^{2}$ in the Einstein frame which depends
on the radial coordinate $\rho $. In the near--horizon region, using the same
symmetry arguments as for the minimally coupled scalar, the operator 
${\mathcal D}$ has the generic form $\star {\mathcal D}=g(\alpha ^{\prime }\Box)$ 
with $g(x)=g_{0}+g_{1}x+\cdots\,$. We then expect the physical solutions
to satisfy 
\begin{equation}
\Box \lambda =\mu ^{2}\lambda \ ,  \label{NearHor}
\end{equation}
where the mass $\mu $ is defined by the relation $g(\alpha ^{\prime }\mu^{2})=0$.

Now we argue that the mass $\mu$ is non--vanishing.
Due to the attractor mechanism, there must a family of extremal solutions
with the same charges $\rho _{n}=\rho _{m}$ and with 
$\lambda \simeq \lambda_{\infty }$ arbitrary at infinity and fixed at the horizon with 
$\lambda_h=0$. To leading order in $\alpha ^{\prime }$, this family
is given by (\ref{hole}) with the harmonic functions $F$, $H$ now given by
\[
F=e^{\lambda _{\infty }}+\frac{\rho _{n}}{\rho }\ ,\ \ \ \ \ \ \
H=e^{-\lambda _{\infty }}+\frac{\rho _{m}}{\rho }\ .
\]
For small $\lambda_{\infty }$ this family of solutions generates fluctuations
of the form studied above. 
Therefore, in the near--horizon region,  we expect that the regular static solution to
(\ref{NearHor})  actually vanishes for 
$\rho\rightarrow 0$. This is only possible if $\mu ^{2}\neq 0$, since in this
case the regular solution is exponentially damped at the horizon. In fact,
when $\mu ^{2}=0$ the regular solution is constant and cannot describe a
solution attracted to $\lambda =0$ at $\rho \rightarrow 0$. In the remainder of this
paper we shall leave
the $\alpha ^{\prime }$ corrected mass $\mu $ as a parameter since its
computation is beyond our reach\footnote{One could naively use the uncorrected 
form of the equation of motion (\ref{uncorrected}) to find 
$\mu ^{2}=8/(\rho _{n}\sqrt{\alpha ^{\prime }})$.
This mass, however, must be corrected with the inclusion of all the higher 
$\alpha ^{\prime }$ terms.}.

We shall see that, for small frequencies, the form of the equations of
motion far from the black hole and near its horizon will be enough to determine the
energy dependence of the cross section.

\subsubsection{Cross section}

We will now analyze the absorption cross section for the s--wave scalar
fluctuation~considered in the last section, in the limit of vanishing
frequency $\omega \rightarrow 0$ of the incoming wave. As it is well known, if
we consider the equations of motion only to leading order in $\alpha
^{\prime }$, the propagation of fluctuations of fixed energy and angular
dependence can be rewritten as an effective one--dimensional Schr\"{o}dinger
problem. Cross sections are then simply related to transmission
probabilities in the presence of an effective potential. In the previous
section we have shown that, to all orders in $\alpha ^{\prime }$ but only in
the asymptotic regions far from the black hole and near its horizon, the
linearized equations for some specific s--wave scalar fluctuation have the
general form 
\[
\Box \Psi =U(\rho )\,\Psi \ 
\]
and are therefore rewritable as effective Schr\"{o}dinger problems. This
asymptotic form of the equations of motion will nonetheless suffice to
determine the energy dependence of the cross section in the limit $\omega\rightarrow 0$. 
Note that we shall \emph{not} assume that the equations of
motion can be recasted in the Schr\"{o}dinger form in the intermediate
region which interpolates between the horizon and Minkowski infinity. The
case of the minimally coupled field $\nu $ corresponds in general to $U=0$.
When considering the fixed scalar $\lambda $ we will, on the other hand,
concentrate on the particular background with $\rho _{n}=\rho _{m}$ and 
$\lambda =0$, $A=C$.

To reduce the equations of motion in the asymptotic regions to a
one--dimensional Schr\"{o}dinger problem, we write the Einstein metric in
the form 
\[
ds_{E}^{\,2}=ds_{2}^{\,2}+r^{2}(x)\, d\Omega_{2}^{2}~.
\]
We have switched radial variable from $\rho $ to the usual Tortoise
coordinate $x=x(\rho )$, which puts the two--dimensional metric in the
conformally flat form 
\[
ds_{2}^{\,2}=a^{2}(x) \left( -dt^{2}+dx^{2}\right) ~.\ 
\]
The horizon is at $x\rightarrow -\infty $ and Minkowski infinity at 
$x\rightarrow +\infty $. Decomposing the field as 
\[
\Psi (t,x,\theta ,\varphi )=\frac{e^{i\omega t}}{r(x)} \,\psi (x)
\]
the asymptotic field equations are in the Schr\"{o}dinger form 
\[
\left( \omega ^{2}+\frac{d^{2}}{dx^{2}}\right) \psi (x)=V(x)\,\psi (x)\ ,
\]
with an effective potential 
\[
V=a^{2}U+\frac{1}{r}\frac{d^{2}}{dx^{2}}\,r\ .
\]

In Appendix A we show that the leading frequency dependence for $\omega
\rightarrow 0$ of the transmission probability $|T|^{2}$ is determined only
by the asymptotic behaviour of the potential, irrespective of the form of the
equations of motion in the intermediate region. A potential tail for 
$x\rightarrow \pm \infty $ such that $x^{2}V(x)$ tends to 
\[
0\ ,\ \ \ \beta ^{2}-1/4\ ,\ \ \ +\infty 
\]
contributes, respectively, to the transmission probability with a factor of 
\[
\omega \ ,\ \ \ \omega ^{2\beta }\ ,\ \ \ \exp \left( \mp \,2\int^{x_{\pm}(\omega)}\sqrt{V(x)}dx\right) \ , 
\]
where $x_{\pm }\left( \omega \right) $ are the classical turning points,
which tend to $\pm \infty $ as $\omega \rightarrow 0$. The numerical
coefficient in front of this leading frequency dependence is determined from
the exact solution of the $\omega =0$ equation with finite amplitude at the
horizon (see Appendix A).

Let us describe in detail the case of the minimally coupled scalar $\nu $.
The following derivation follows closely \cite{Das} but shows that the
result can be obtained without the knowledge of the equations of motion in
the intermediate region. We start with a purely infalling wave at the
horizon 
\[
\psi (x)\simeq e^{i\omega x}\ ,\ \ \ \ \ \ \ \ x\rightarrow -\infty \ .
\]
Since the effective potential at the horizon decays faster than $1/x^{2}$,
the asymptotic wave length grows faster than $|x_{-}|$ as $\omega $ tends to
zero. Therefore, the wave function will penetrate far beyond the turning
point keeping its profile, i.e. 
\begin{equation}
\psi (x)\simeq 1+i\omega x\ ,\ \ \ \ \ \ \ \ x_{-}\ll x\ll \epsilon x_{-}\ ,
\label{asym}
\end{equation}
where $\epsilon \ll 1$. In the region $x_{-}\ll x\ll x_{+}$ we have $\omega
^{2}\ll V(x)$ so that we can drop the $\omega ^{2}$ term in Schr\"{o}dinger
equation and consider the solution at zero frequency. In fact, the equations
of motion need not even be of the Schr\"{o}dinger form in the intermediate
region. We only need the solution at $\omega =0$ which connects to the
asymptotic form of the field (\ref{asym}). Moreover, since we are interested
in the $\omega \rightarrow 0$ limit, we only need to evolve the constant
term $1$. For this purpose, we use the constant solution for $\nu $ which is
valid to all orders in $\alpha ^{\prime }$.
At infinity, the effective action for s--wave perturbations of the field $\nu$ takes the form
$$
S\simeq \frac{5}{4 G_N} \int dt dx \left[ (\partial_t \psi)^2  -  (\partial_x \psi)^2  
-\frac{\partial_x^2 r}{r} \psi^2  \right]\ ,
$$
with $\psi(t,x)=\nu(t,x) \,r(x)$.
As explained above, symmetry arguments fix  the effective action in the near--horizon region to have the form
(\ref{nhaction}), yielding
$$
S\simeq \frac{5}{4 G_N} \int dt dx \, \frac{c}{2}\, R_h^{\,2} \Big[ (\partial_t \nu)^2  -  (\partial_x \nu)^2  \Big]\ ,
$$
where $R_{h}$ is the horizon radius in the Einstein frame and
$c/2$ is the proportionality constant between the effective near--horizon metric $K_{\alpha \beta}$
seen by the scalar field and the string metric.
Thus, the exact solution $\nu=\sqrt{2}/(\sqrt{c} R_h)$ corresponds to $\psi=1$ at the horizon and to 
$\psi\simeq\sqrt{2}x/(\sqrt{c} R_h)$ at infinity.
We can then evolve the term $1$ in equation (\ref{asym}), obtaining 
\[
\psi (x)\simeq \sqrt{\frac{2}{c}}\frac{x}{R_{h}}\ ,\ \ \ \ \ \ \ \ \epsilon x_{+}\ll x\ll x_{+}\ ,
\]
and, since once again one has a fast decaying potential at infinity, 
\[
\psi (x)\simeq \frac{1}{i \sqrt{2c} \, \omega  R_{h}}\left( e^{i\omega x}-e^{-i\omega
x}\right) \ ,\ \ \ \ \ \ \ \ x\rightarrow \infty \ .
\]
Hence, the transmission probability for the s--wave is 
\[
|T_{0}|^{2}\simeq 2 c R_{h}^{\,2}\omega ^{2}\ .
\]
Moreover, since the total cross section is given by 
\[
\sigma _{abs}=\frac{\pi }{\omega ^{2}}\sum_{l=0}^{\infty }(2l+1)|T_{l}|^{2}\
,
\]
and for small frequencies only the s--wave contributes, we obtain 
\[
\sigma _{abs}\simeq \frac{c}{2} A_{h} \simeq c\, G_N S \ .
\]
The result  $\sigma _{abs}\simeq  A_{h} $ is generic for fields obeying the massless Klein--Gordon
equation throughout the whole space \cite{Das}. Here we have generalized it to the case where we only
know the interpolating constant solution, together with the fact
that the field obeys the massless Klein--Gordon equation in both asymptotic
regions.

Let us now consider the case of the fixed scalar $\lambda $. At infinity the
potential decays as $1/x^{3}$, while near the horizon it goes as 
\[
V\simeq \frac{\mu ^{2}R_{h}^{\,2}}{x^{2}}\ ,
\]
where $\mu $ is the unknown mass term. In this case the small frequency
behaviour of the absorption cross section reads 
\[
\sigma _{abs}\sim \omega ^{2\beta -1}\ ,
\]%
where 
\[
2\beta =\sqrt{1+(2\mu R_{h})^{2}}~.
\]%
As explained in the previous section, \emph{the attractor mechanism implies
that} $\mu >0$, so that the absorption cross section for the fixed scalar
vanishes in the $\omega \rightarrow 0$ limit\footnote{If one naively neglects 
the $\alpha ^{\prime }$ corrections to the mass $\mu 
$ (see previous footnote) the cross section satisfies $\sigma _{abs}\sim
\omega ^{\sqrt{17}-1}\sim \omega ^{3.12}$. }. On the other hand, we cannot
fix the normalization since we do not have an explicit zero--frequency
solution valid in the entire intermediate region. The absence of absorption
at zero frequency comes as no surprise since this scalar has its value at
the horizon fixed by the hole charges. This suppresses fluctuations close to
the horizon and therefore the associated infalling flux.

\section{String phase}

In this section we use perturbative string theory to study the string states
corresponding to the black hole of the previous section and to compute their
absorption cross section. Let $X_{L}^{\mu }$ ($\mu =0,\cdots ,25)$ and 
$X_{R}^{\mu },~\tilde{\psi}^{\mu }~$($\mu =0,\cdots ,9$) be, respectively,
the left-- and right--moving worldsheet matter fields of the Heterotic string,
where directions $\mu =0,\cdots ,3$ correspond to the four--dimensional
spacetime, directions $\mu =4,5,\cdots ,9$ are compactified on $S^{1}\times
T^{5}$ and the remaining left moving directions $\mu =10,\cdots ,25$
correspond to the internal compact Heterotic torus. The matter Virasoro
constraints read $L_{0}=\tilde{L}_{0}=0$, with
\[
\begin{array}{l}
\displaystyle{L_{0}=N+\frac{\alpha ^{\prime }}{4}\,{\bf p}_{L}^{\,2}-1}\,,
\spa{0.5}\\
\displaystyle{\tilde{L}_{0}=\tilde{N}+
\frac{\alpha^{\prime}}{4}\,{\bf p}_{R}^{\,2}-\frac{1}{2}\,\delta _{NS}\,,}
\end{array}
\]
where $\delta _{NS}=1$ in the NS sector and vanishes in the R sector. The $N$'s 
and ${{\bf p}}$'s are the usual left-- and right--moving levels and
momenta.

\subsection{String states}

The black hole is a BPS massive state of the Heterotic theory compactified
on $S^{1}\times T^{5}$. The corresponding string has Ka\l u\.{z}a--Klein momentum
and winding $(n,m)$ along the $S^{1}$ direction, such that 
\[
p_{L,R}^{\,4}=\left( \frac{n}{R}\mp \frac{mR}{\alpha ^{\prime }}\right) \ . 
\]
The $T^{5}$ directions are mere spectators and we shall consider black holes
which are chargeless under the internal gauge group, so that $p_{L,R}^{\,\mu}=0$ 
for $\mu \geq 5$. Therefore, the string has total left/right momentum 
\[
{\bf p}_{L,R}=(p,p_{L,R}^{\,4},0)~. 
\]
In this notation, the unbold momentum $p$ is four--dimensional. The BPS
condition
\[
{\bf p}_{R}\cdot {\bf p}_{R}=0 
\]
fixes the four--dimensional mass to be 
\[
M^{2}=-p\cdot p=\left( \frac{n}{R}+\frac{mR}{\alpha ^{\prime }}\right) ^{2}\ . 
\]
The Virasoro constraints $L_{0}=\tilde{L}_{0}=0$ then imply 
\[
N=nm+1\,,\ \ \ \ \ \ \ \ \ \ \ \ \tilde{N}=\frac{1}{2}\,\delta _{NS}~. 
\]
There are many possible ways of satisfying these conditions corresponding to
different black hole microstates. In the string phase we shall average over
all the string microstates.

In the NS sector the string states are space--time bosons with vertex
operator in the $-1$ picture given by 
\begin{equation}
{{\mathcal V}}_{i}(z,\bar{z})=\,c\,\tilde{c}\,e^{-\tilde{\phi}}~\,{{\mathcal V}
}_{i}(z)\ \boldsymbol{ \zeta}\cdot  \boldsymbol{ \tilde{\psi}}\,e^{\,i\,{\bf p}
_{R}\cdot {\bf X}_{R}},  \label{vertex}
\end{equation}
where the holomorphic matter part ${{\mathcal V}}_{i}(z)$ is a conformal
primary with $L_{0}=0$ and momentum ${\bf p}_{L}$. The string
polarization vector $ \boldsymbol{ \zeta}$ satisfies the physical condition 
\[
 \boldsymbol{ \zeta }\cdot {\bf p}_{R}=0\ ,  \label{BHpolar}
\]
with normalization $ \boldsymbol{ \zeta }\cdot \boldsymbol{ \zeta }=1$.

In the R sector the string states are space--time fermions with vertex
operator in the $-1/2$ picture given by 
\begin{equation}
{{\mathcal V}}_{i}(z,\bar{z})=\,c\,\tilde{c}\,e^{-\tilde{\phi}/2}\,\,{{\mathcal V}}_{i}(z)\ 
u^{\alpha }\tilde{\Theta}_{\alpha }\,e^{\,i\,{\bf p}_{R}\cdot {\bf X}_{R}},  \label{fermiBH}
\end{equation}
where $u^{\alpha }$ is a chiral spinor and $\tilde{\Theta}_{\alpha }$ is the
R state spin field. Physical states obey the Dirac equation $\sp u=0$ and
are normalized by $\bar{u}\Gamma ^{0}u=2p^{0}$.

For fixed charges $(n,m)$ there are many possible left--moving string
microstates which are conveniently parametrized using DDF operators \cite{DelGiudice:1971fp}. 
To be specific, we work in the center--of--mass frame,
so that the four--dimensional momentum $p$ is given by
\[
p=\left( M,0,0,0\right) ~.
\]
Choosing ${\bf k}$ to be the null vector with $k^{\mu }=0$ for $\mu \geq 4
$ and
\[
k=-\frac{2}{M\alpha ^{\prime }}\left( 1,1,0,0\right)~,
\]%
we define the DDF operators 
\[
A_{n}^{i}\equiv \sqrt{\frac{2}{\alpha ^{\prime }}}\,\,\oint \frac{dz}{2\pi i}
\,i\partial X_{L}^{i}\,e^{\,i\,n\,{\bf k}\cdot {\bf X}_{L}}(z)\ ,
\]%
with $n$ integer and $X^{i}$ $(i=2,\cdots ,25)$ one of the 24 directions
transverse to ${\bf k}$. These operators obey 
\[
\lbrack L_{m},A_{n}^{i}]=0\ ,\ \ \ \ \ \ \ \ \ \ \ \
[A_{n}^{i},A_{m}^{j}]=n\,\delta ^{ij}\,\delta _{m+n}\ ,
\]
when acting on states whose momentum ${\bf p}_{L}$ satisfies 
$\alpha^{\prime }{\bf p}_{L}\cdot {\bf k}=2$. Then, the normalized
holomorphic matter part of the vertex operator can be written as 
\[
{{\mathcal V}}_{i}(z)=\prod_{i=2}^{25}\prod_{n=1}^{\infty }
\frac{1}{\sqrt{r_{n}^{i}!\,n^{r_{n}^{i}}}}\left( A_{-n}^{i}\right)
^{r_{n}^{i}}e^{\,i\,\left( {\bf p}_{L}+N{\bf k}\right) \cdot {\bf X}_{L}(z)}\ ,
\]
where the $r_{n}^{i}$ are occupation numbers satisfying $\sum_{i=2}^{25}%
\sum_{n=1}^{\infty }n\,r_{n}^{i}=N$. In the operator language the string
states are constructed by applying DDF operators to the left--moving
tachyonic ground state with momentum ${\bf p}_{L}+N{\bf k}$. The
string left--moving degeneracy is then given by the number of partitions $%
\{r_{n}^{i}\}$ of the level $N$ in integers of 24 colours, corresponding to
the 24 transverse directions.

\subsection{Absorption amplitude}

\begin{figure}[tbp]
\centering\includegraphics[width=8cm]{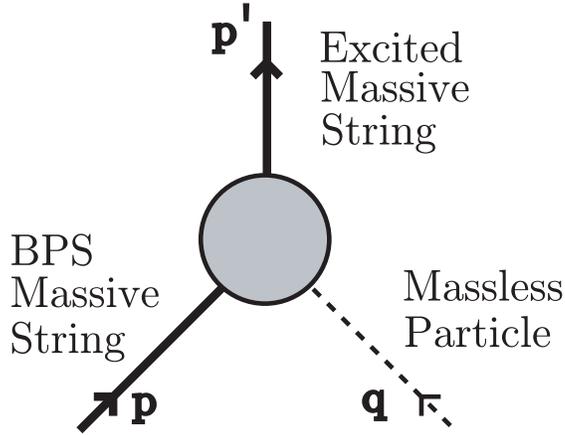} 
\caption{{\protect\small {A BPS massive string state absorbs a massless
particle becoming a low lying excited string state above the BPS bound.}}}
\label{fig3}
\end{figure}

In this section we will determine the absorption cross section of a massless
field by the massive BPS string states just described as represented in Figure 
\ref{fig3}. We compute
first the absorption amplitude of a massless boson by bosonic microstates of
the BPS massive string. The incoming massive state has vertex operator given
by (\ref{vertex}). The external massless particle is described by the vertex
operator in the $0$ picture 
\[
{{\mathcal V}}(z,\bar{z})=\,\frac{2}{\alpha ^{\prime }}\,c\,\tilde{c}
\,i\partial X_{L}^{l}\,\left( i\,\bar{\partial}X_{R}^{j}+\frac{\alpha^{\prime }}{2}\,
{\bf q}\cdot \boldsymbol{\tilde{\psi}}\,\,\tilde{\psi}^{j}\right) 
e^{\,i{\bf q}\cdot \left( {\bf X}_{L}+{\bf X}_{R}\right) }\ ,
\]%
where ${\bf q}$ is the particle's momentum, with the only non--vanishing
components given by $q^{0}=q^{1}=\omega $, where $\omega $ is the energy 
in the frame of the BPS string state. The transverse
polarization directions $l$ and $j$ then run from $2$ to $25$ and from $2$
to $10$, respectively. Note that we have chosen, for reasons that will
become clear soon, the momentum ${\bf k}$ used to define the DDF operators
to be collinear with the momentum ${\bf q}$, so that
\[
{\bf q=-}\varepsilon {\bf k~,}~\ \ \ \ \ \ \ \ \ \ \ \ \ \ \ \ \ \ \ \
\varepsilon =\frac{\omega M\alpha ^{\prime }}{2}\ .
\]
The excited massive string final state will have momentum ${\bf p}%
_{L,R}^{\prime }={\bf p}_{L,R}+{\bf q}$ and vertex operator of the
form 
\[
{{\mathcal V}}_{f}(z,\bar{z})=\,c\,\tilde{c}\,e^{-\tilde{\phi}}\,\,
{{\mathcal V}}_{f}(z)\ \tilde{{\mathcal V}}_{f}(\bar{z}),
\]
where ${{\mathcal V}}_{f}(z)$ and $\tilde{{\mathcal V}}_{f}(\bar{z})$ are
conformal primaries of the matter theory with $L_{0}=\tilde{L}_{0}=0$.

The tree level $S$--matrix element for this process is given by 
\[
\frac{4\kappa \,}{\alpha ^{\prime }}\left\langle {{\mathcal V}}_{f}^{\ast}(z_{3},\bar{z}_{3})\,
{{\mathcal V}}(z_{2},\bar{z}_{2}){{\mathcal V}}_{i}(z_{1},\bar{z}_{1})\,\right\rangle =
\,i\,(2\pi )^{4}\,\delta^{(4)}(p'-p-q)\,\,{\mathcal M}\ .
\]
The normalization is given by the four--dimensional gravitational coupling $\kappa$ 
defined as usual by 
\[
2\kappa^{2}=16\pi G_N = 2\pi g^2\alpha'
\]
and is fixed, for example, by matching the three graviton coupling with the
one coming from the low energy effective action. The matrix element ${\mathcal M}$ is then given by
\begin{equation}
{\mathcal M} =\frac{4\kappa}{\alpha^{\prime }}\,C\tilde{C}~,
\label{matrixEl}
\end{equation}
where the dimensionless constants $C$ and $\tilde{C}$ come, respectively,
from the holomorphic and anti--holomorphic parts of the correlation function 
\[
\begin{array}{l}
\displaystyle{C=\sqrt{\frac{2}{\alpha^{\prime}}}\,{z}_{12}{z}_{23}{z}_{13}
\left\langle {{\mathcal V}}_{f}^{\ast }({z}_{3})~i{\partial }X_{L}^{l}\,
e^{\,i{\bf q}\cdot {\bf X}_{L}}({z}_{2})~{{\mathcal V}}_{i}({z}_{1})\right\rangle ~,} 
\spa{0.5}\\ 
\displaystyle{ \tilde{C}=\,\sqrt{\frac{2}{\alpha^{\prime}}}\,\bar{z}_{12}\bar{z}_{23}
\left\langle \tilde{{\mathcal V}}_{f}^{\ast }(\bar{z}_{3})
\left(i\,\bar{\partial}X_{R}^{j}+\frac{\alpha ^{\prime }}{2}\,
{\bf q}\cdot \boldsymbol{\tilde{\psi}}\,\,\tilde{\psi}^{j}\right) \!e^{\,i{\bf q}\cdot {\bf X}
_{R}}(\bar{z}_{2}) \boldsymbol{\zeta } \cdot \boldsymbol{\tilde{\psi}}\,e^{\,i\,{\bf p}
_{R}\cdot {\bf X}_{R}}(\bar{z}_{1})\right\rangle }\ ,
\end{array}
\]
where $z_{ij}=z_{i}-z_{j}$. The $z_{i}$ dependence of the matter 3--pt
functions is fixed by conformal invariance and precisely cancels the
explicit factors of $z_{ij}$ coming from the ghosts. Thus, one can take the
limit $z_{2}\rightarrow z_{1}$ and replace the BPS massive string and
massless particle vertex operators by their OPE. Non--singular terms of the
OPE disappear in the $z_{2}\rightarrow z_{1}$ limit because of the explicit
factors of ${z}_{12}$ and $\bar{z}_{12}$. On the other hand, terms more
singular than ${z}_{12}^{\,-1}$ can not contribute because they would give a
divergent answer. Hence, the only contribution comes from the ${z}%
_{12}^{\,-1}$ term in the OPE, which is given by the contour integral of the
massless particle vertex around the massive string one. The result is a
conformal primary with the same weight of the massive string vertex because
the contour integral of weight $1$ holomorphic primaries defines an operator
that commutes with the Virasoro generators. One is then left with a
two--point function, so that this conformal primary is, up to normalization,
the final excited string state 
\[
\begin{array}{l}
\displaystyle{C\,{{\mathcal V}}_{f}({z}_{1})=\sqrt{\frac{2}{\alpha ^{\prime
}}}\oint_{z_1} \frac{dz_{2}}{2\pi i}\,\,i{\partial }X_{L}^{l}\,e^{\,i{\bf q}
\cdot {\bf X}_{L}}({z}_{2})\,{{\mathcal V}}_{i}({z}_{1})~,} 
\spa{0.5}\\ 
\displaystyle{\tilde{C}\,\tilde{{\mathcal V}}_{f}(\bar{z}_{1})=\sqrt{\frac{2
}{\alpha ^{\prime }}}\oint_{\bar{z}_1} \frac{d\bar{z}_{2}}{2\pi i}\left( i\,\bar{\partial
}X_{R}^{j}+\frac{\alpha ^{\prime }}{2}\,{\bf q}\cdot \boldsymbol{ \tilde{\psi}}\,\,
\tilde{\psi}^{j}\right) e^{\,i{\bf q}\cdot {\bf X}_{R}}(\bar{z}_{2})~\,
 \boldsymbol{\zeta }\cdot  \boldsymbol{\tilde{\psi}}\,e^{\,i\,{\bf p}\cdot {\bf X}_{R}}(
\bar{z}_{1})}\ .
\end{array}
\]

The only quantities that remain to be computed are the above normalization
constants $C$ and $\tilde{C}$. Let us start by computing $C$. The
computation greatly simplifies since we have chosen ${\bf k}$, appearing
in the DDF operators, to be collinear with the momentum ${\bf q=-}%
\varepsilon {\bf k}$ of the massless particle. Using $\alpha ^{\prime }%
{\bf p}\cdot {\bf k}=2$, it is easy to show that $\varepsilon =\omega
M\alpha ^{\prime }/2$ is related to the square of the center--of--mass
energy $s=-(p+q)^{2}$ through 
\[
\varepsilon =\frac{\alpha ^{\prime }}{4}\left( s-M^{2}\right) \ .
\]%
The absorption process can only occur if $s=M^{\prime 2}$, for some excited
level of the Heterotic string with mass $M^{\prime }$ and the same charges $n
$ and $m$. Therefore, 
\begin{equation}
\varepsilon =\frac{\alpha ^{\prime }}{4}\left( M^{\prime 2}-M^{2}\right) \ 
\label{kinematics}
\end{equation}%
is bound to be an integer by the $L_{0}$ and $\tilde{L}_{0}$ physical
conditions 
\[
\varepsilon =\tilde{N}^{\prime }-\frac{1}{2}\,\delta _{NS}=N^{\prime }-N\ ,
\]%
where $N^{\prime }$ and $\tilde{N}^{\prime }$ are the levels of the final
state. Finally, using the DDF operator parametrization of the string
microstate ${{{\mathcal V}}_{i}({z})}$ one obtains 
\[
C\,{{\mathcal V}}_{f}({z})=A_{-\varepsilon
}^{l}\prod_{i=2}^{25}\prod_{n=1}^{\infty }\frac{1}{\sqrt{r_{n}^{i}!
\,n^{r_{n}^{i}}}}\left( A_{-n}^{i}\right) ^{r_{n}^{i}}e^{\,i\,\left( {\bf 
p}_{L}+N{\bf k}\right) \cdot {\bf X}_{L}(z)}\ ,
\]
which fixes 
\[
C=\sqrt{\varepsilon (r_{\varepsilon }^{l}+1)}~.
\]
To determine the other constant $\tilde{C}$ we define the DDF operator 
\[
\tilde{A}_{-\varepsilon }^{j}\equiv \sqrt{\frac{2}{\alpha ^{\prime }}}\oint 
\frac{d\bar{z}}{2\pi i}{\left( i\,\bar{\partial}X_{R}^{j}+\frac{\alpha^{\prime }}{2}\,
{\bf q}\cdot  \boldsymbol{ \tilde{\psi}}\,\,\tilde{\psi}^{j}\right) 
}\,e^{\,i{\bf q\cdot X}_{R}}(\bar{z})\ ,
\]
for ${\bf q=-}\varepsilon {\bf k}$ and integral $\varepsilon $, and
write 
\[
\tilde{C}\,\tilde{|{\mathcal V}}_{f}\rangle =\tilde{A}_{-\varepsilon }^{j}|
\tilde{{\mathcal V}}_{i}\rangle \ ,\ ~\ \ \ \ \ \ \ \ \ \ \ \ \tilde{C}
^{2}=\langle \tilde{{\mathcal V}}_{i}|\tilde{A}_{\varepsilon }^{j}\tilde{A}
_{-\varepsilon }^{j}|\tilde{{\mathcal V}}_{i}\rangle \ ,
\]
where $|\tilde{{\mathcal V}}_{i}\rangle $ denotes the right--moving matter
part of the initial string state. Then, since $[\tilde{A}_{\varepsilon }^{j},
\tilde{A}_{-\varepsilon }^{j}]=\varepsilon $ and since 
$\tilde{A}_{\varepsilon }^{j}|\tilde{{\mathcal V}}_{i}\rangle =0$ for $\varepsilon \geq 1$, 
we arrive at the simple result 
\[
\tilde{C}=\sqrt{\varepsilon }~.
\]
We have then computed the absorption matrix element (\ref{matrixEl}) 
\begin{equation}
{{\mathcal M}}=\frac{4\kappa }{\alpha ^{\prime }}\,\varepsilon \sqrt{r_{\varepsilon }^{l}+1}\ .  
\label{Mabs}
\end{equation}

\subsubsection{Fermionic states}

The computation of the absorption amplitude of massless bosons by fermionic
microstates of the BPS massive string proceeds along similar lines, with the
same result (\ref{Mabs}) for the absorption amplitude. The
initial string vertex operator is given by (\ref{fermiBH}) and the final one
can be taken to be ${{\mathcal V}}_{f}(z,\bar{z})=\,c\,\tilde{c}\,
e^{-\tilde{\phi}/2}\,\,{{\mathcal V}}_{f}(z)\ \tilde{{\mathcal V}}_{f}(\bar{z})$, where 
${{\mathcal V}}_{f}(z)$ and $\tilde{{\mathcal V}}_{f}(\bar{z})$ are conformal
primaries of the matter theory with $L_{0}=\tilde{L}_{0}=0$ and with 
$\tilde{{\mathcal V}}_{f}$ in the Ramond sector. One then needs to write the massless
particle vertex in the $-1$ picture, 
\[
{{\mathcal V}}(z,\bar{z})=\sqrt{\frac{2}{\alpha ^{\prime }}}\,c\,\tilde{c}
\,e^{-\tilde{\phi}}\,i\partial X_{L}^{l}~\tilde{\psi}^{j}\,e^{\,i{\bf q}
\cdot \left( {\bf X}_{L}+{\bf X}_{R}\right) }\ .
\]
The holomorphic part of the correlator is identical, giving 
$C=\sqrt{\varepsilon (r_{\varepsilon }^{l}+1)}$. The
antiholomorphic part reads
\[
{\tilde{C}=\,\bar{z}_{12}\bar{z}_{23}\bar{z}_{13}\left\langle e^{-\tilde{\phi}/2}
\tilde{{\mathcal V}}_{f}^{\ast }(\bar{z}_{3})\,\,e^{-\tilde{\phi}}
\tilde{\psi}^{j}e^{\,i{\bf q}\cdot {\bf X}_{R}}(\bar{z}_{2})\,\,
e^{-\tilde{\phi}/2}u^{\alpha }\tilde{\Theta}_{\alpha }\,
e^{\,i\,{\bf p}_{R}\cdot {\bf X}_{R}}\!\!~(\bar{z}_{1})\right\rangle ,}
\]
with the relevant DDF operator now given by
\[
\tilde{A}_{-\varepsilon }^{j}\equiv \oint \frac{d\bar{z}}{2\pi i}\,
e^{-\tilde{\phi}}\,\tilde{\psi}^{j}\,e^{\,i{\bf q\cdot X}_{R}}(\bar{z})
\]%
for ${\bf q=-}\varepsilon {\bf k}$ and integral $\varepsilon $. As
before, $\tilde{A}_{\varepsilon }^{j}$ annihilates the initial state and $[%
\tilde{A}_{\varepsilon }^{j},\tilde{A}_{-\varepsilon }^{j}]=$ $\varepsilon $
determines the constant $\tilde{C}=\sqrt{\varepsilon }$.

\subsection{Absorption cross section}

The absorption cross section is given by 
\[
\sigma _{abs}=\frac{\pi }{4M\omega (M+\omega )}\sum_{M^{\prime }}\delta
\left( \sqrt{M^{\prime 2}+\omega ^{2}}-M-\omega \right) \left\vert {{\mathcal 
M}}\right\vert ^{2}\ , 
\]%
where $\omega $ is the massless particle energy in the massive string
reference frame and the amplitude ${{\mathcal M}}$ is given by (\ref{Mabs}).
Using the kinematical relation (\ref{kinematics}) one can simplify the
absorption cross section expression to 
\[
\sigma _{abs}=\frac{\pi \alpha ^{\prime }}{8M\omega }\sum_{\varepsilon
}\delta \left( \frac{\alpha ^{\prime }}{2}M\omega -\varepsilon \right)
\left\vert {{\mathcal M}}\right\vert ^{2}\ , 
\]%
so that, for a given microstate, one has 
\[
\sigma _{abs}=\pi \kappa ^{2}\sum_{\varepsilon }\varepsilon \left(
r_{\varepsilon }^{l}+1\right) \delta \left( \frac{\alpha ^{\prime }}{2}%
M\omega -\varepsilon \right) \ . 
\]

The black hole should be thought of as a classical mixture of all its
microstates. In Appendix B we show that 
\[
\langle r_{\varepsilon }^{l}\rangle =\frac{1}{\exp\left(\frac{2\pi \varepsilon }{\sqrt{N}}\right)-1}\ , 
\]
for $N\gg 1$ and where the average $\langle \ \rangle $ is taken over all
left--moving microstates of the BPS string with charges $n$ and $m$. Thus,
recalling that $2\kappa ^{2}=2\pi g^{2}\alpha ^{\prime }$, we have
\[
\langle \sigma _{abs}\rangle =\pi ^{2}g^{2}\alpha ^{\prime
}\sum_{\varepsilon }\frac{\varepsilon }{1-\exp \left( -{\frac{2\pi
\varepsilon }{\sqrt{N}}}\right) }\,\,\delta \left( \frac{\alpha ^{\prime }}{2}
M\omega -\varepsilon \right) ~. 
\]
For energies much larger than the black hole mass gap,
\[
\omega \gg \frac{2}{\alpha ^{\prime }M}\ , 
\]
we can drop the sum of delta functions. Notice that this condition always holds 
in the classical limit, since  $\alpha' M\rightarrow \infty$ and therefore 
the mass gap vanishes\footnote{There are other excitations of the BPS string,
particularly those associated to the addition of $NS5/\overline{NS5}$ brane pairs \cite{Mathur:2005ai}.
Although for $g^2\sqrt{N}\gg 1$ the mass gap for these excitations 
is smaller than the fundamental string excitations mass gap, $2/(\alpha'M)$, these are entropically disfavoured provided the
energy $\omega$ is kept finite.}.
The last expression can then be recasted in the form 
\[
\langle \sigma _{abs}\rangle =\frac{\pi }{2}\,g^{2}\,\alpha ^{\prime }\sqrt{N%
}\,\frac{\omega }{2T_{L}}\,\frac{1}{1-\exp \left( -\frac{\omega }{2T_{L}}%
\right) }\ , 
\]%
where 
\[
T_{L}=\frac{\sqrt{N}}{2\pi \alpha ^{\prime }M} 
\]%
is the effective left--moving temperature. This result is universal in the sense that it
does {\em not} depend on the massless particle polarization.

To compare with the gravity phase we must consider the low energy limit, $%
\omega \ll T_{L}$, where 
\[
\langle \sigma _{abs}\rangle \simeq \frac{\pi }{2}\,g^{2}\,\alpha ^{\prime }\sqrt{N}
\simeq G_N S \ .
\]%
This result clearly indicates the difference between string and black hole
phases. Indeed, we saw in the gravity phase that the minimally coupled
scalars and the fixed scalars absorb quite differently. In particular, in
the limit $\omega \rightarrow 0$, the fixed scalars do not absorb since
their values are fixed at the horizon. A finite cross section for {\em all}
particle polarizations signals therefore the absence of a horizon for small
values of the coupling.
Figure \ref{fig2} of the Introduction schematically summarizes this discussion.

\subsection{String size}

In this last section, we estimate the size of the BPS massive string state
as a function of the string coupling, following \cite{Damour}\footnote{The main differences are
that here we consider BPS states and that, by parameterizing the states with DDF
operators, we have full control of the Virasoro constraints.}. Having chosen
the center--of--mass frame for the initial state with $p^{\mu }=0$ for $\mu
=1,2,3$, we can define the size ${\mathcal R}$ of a given microstate $|{%
{\mathcal V}}_{i}\rangle $ via 
\[
{\mathcal R}^{2}\equiv \langle {{\mathcal V}}_{i}|\left( X^{\mu }-X_{CM}^{\mu
}\right) ^{2}|{{\mathcal V}}_{i}\rangle =\alpha ^{\prime }\sum_{m=1}^{\infty }
\frac{1}{m^{2}}\,\langle {{\mathcal V}}_{i}|\alpha _{-m}^{\mu }\alpha _{m}^{\mu
}+\tilde{\alpha}_{-m}^{\mu }\tilde{\alpha}_{m}^{\mu }|{{\mathcal V}}
_{i}\rangle +\alpha ^{\prime }\sum_{m=1}^{\infty }\frac{1}{m}\ ,
\]
where $X_{CM}^{\mu }$ are the zero modes of the string and where
\[
\alpha _{m}^{\mu }=\sqrt{\frac{2}{\alpha ^{\prime }}}\oint \frac{dz}{2\pi }
\,z^{m}\,\partial X_{L}^{\mu }(z)
\]
are the usual raising and lowering operators. The physical meaning of the
divergent sum, which comes from normal ordering the operators, is discussed
in \cite{Karliner}. This is a state independent  effect that we
shall discard. The BPS states under consideration do not have right--moving
excitations so that the $\tilde{\alpha}$ contribution to ${\mathcal R}^{2}$
is zero. Then, we have 
\begin{equation}
{\mathcal R}^{2}=\alpha ^{\prime }\sum_{m=1}^{\infty }\frac{1}{m^{2}}\,\Delta
_{m}\ ,  \label{RWsize}
\end{equation}%
where $\Delta _{m}\equiv \langle {{\mathcal V}}_{i}|\alpha _{-m}^{\mu }\alpha
_{m}^{\mu }|{{\mathcal V}}_{i}\rangle $.

The parametrization of the microstates ${{\mathcal V}}_{i}$ using DDF
operators explained in Section 3.1 breaks rotational invariance
because one has to select a spatial direction for the null vector ${\bf k}
$. Nevertheless, after averaging over all possible microstates the rotational
symmetry should be recovered. Therefore, for simplicity, we compute the
average size of the string in a direction $X^{\mu }$ $(\mu =2,3)$ orthogonal
to ${\bf k}$. In Appendix C we show explicitly that the same result holds
if we choose the spatial direction of the null vector ${\bf k}$. For each
left--moving microstate 
\[
|{{\mathcal V}}_{i}\rangle =\prod_{i=2}^{25}\prod_{n=1}^{\infty }\frac{1}{\sqrt{%
r_{n}^{i}!\,n^{r_{n}^{i}}}}\left( A_{-n}^{i}\right) ^{r_{n}^{i}}|{\bf p}%
_{L}+N{\bf k}\rangle \ ,
\]%
we compute $\Delta _{m}$ by using the commutation relations 
\[
\lbrack \alpha _{m}^{\mu },A_{n}^{i}]=m\delta ^{\mu i}B_{m}^{n}\ ,\ \ \ \ \
\ \ \ \ \ \ \ \ \ [A_{n}^{i},B_{m}^{l}]=[B_{m}^{n},B_{p}^{l}]=0\ ,
\]%
where 
\[
B_{m}^{n}=\oint \frac{dz}{2\pi i}\,z^{m-1}\,e^{\,i\,n\,{\bf k}\cdot 
{\bf X}_{L}(z)}\ .
\]%
We obtain the expression
\[
\Delta _{m}=\left\vert \sum_{l=1}^{\infty }\frac{mr_{l}^{\mu }}{\sqrt{%
r_{l}^{\mu }!\,l^{r_{l}^{\mu }}}}\left( A_{-l}^{\mu }\right) ^{r_{l}^{\mu
}-1}\prod_{n\neq l}\frac{1}{\sqrt{r_{n}^{\mu }!\,n^{r_{n}^{\mu }}}}\left(
A_{-n}^{\mu }\right) ^{r_{n}^{\mu }}B_{m}^{-l}|{\bf p}_{L}+N{\bf k}%
\rangle \right\vert ^{2}\ .
\]%
Since $B_{m}^{-l}|{\bf p}_{L}+N{\bf k}\rangle =0$ for $l<m$ and $%
\left( B_{m}^{-l}\right) ^{\dagger }=B_{-m}^{l}$ the sum over $l$ reduces to
the single term $l=m$. Finally, using $B_{m}^{-m}|{\bf p}_{L}+N{\bf k}%
\rangle =|{\bf p}_{L}+(N-m){\bf k}\rangle $, we obtain the intuitive
result
\[
\Delta _{m}=mr_{m}^{\mu }\ .
\]
Averaging over the microstates and in the large $N$ limit (see Appendix B)
we have 
\[
\langle \Delta _{m}\rangle =\frac{m}{\exp {\frac{2\pi m}{\sqrt{N}}}-1}\ .
\]
The sum in (\ref{RWsize}) is dominated by small values of $m$. Thus we only
need 
\[
\langle \Delta _{m}\rangle \simeq \frac{\sqrt{N}}{2\pi }\ ,
\]
for $m\ll \sqrt{N}$, obtaining 
\[
\langle {\mathcal R}^{2}\rangle \simeq \frac{\pi }{12}\,\alpha ^{\prime }%
\sqrt{N}\ ,
\]
for large $N$.

Besides the average string size one might want to know how many string
states have a given size. This size distribution is described by an entropy
function $S_{0}(N,{\mathcal R})$. Following \cite{Damour} we show in Appendix
B that, in the large $N$ limit, 
\[
S_{0}\left( N,{\mathcal R}\right) \simeq 4\pi \sqrt{N}\left( 1-
\frac{\alpha ^{\prime }}{32{\mathcal R}^{2}}\right) \ ,\ \ \ \ \ \ \ \ \ \ \ \
\ \alpha ^{\prime }\ll {\mathcal R}^{2}\ll \alpha ^{\prime }\sqrt{N}~,
\]
and 
\[
S_{0}\left( N,{\mathcal R}\right) \simeq 4\pi \sqrt{N}\left( 1-\frac{{\mathcal 
R}^{2}}{2\alpha ^{\prime }N}\right) \ ,\ \ \ \ \ \ \ \ \ \ \ \ \alpha ^{\prime }
\sqrt{N}\ll {\mathcal R}^{2}\ll \alpha ^{\prime }N\ .
\]
These results suggest a description of the free string as a random walk with 
$\sqrt{N}$ steps and step size of order $\sqrt{\alpha ^{\prime }}$.

\subsubsection{Self--gravitating string}

We would like to estimate the variation of the string size when the string
coupling is increased. We have just seen that at zero coupling the string
is, on average, very large (${\mathcal R}^{2}\sim \alpha ^{\prime }\sqrt{N}$)
and can be described as a random polymer. On the other hand, for large
enough $g$, we expect the string to collapse into the black hole described in
Section 2, whose size is of order $\sqrt{\alpha ^{\prime }}$. Therefore, as
we vary the coupling, we expect the string to contract due to gravitational
interactions.

To study the contracting string for small $g$ we shall use the polymer
picture of a low interacting string \cite{Khuri}. This model was used to
describe non--BPS string states, which, as the string contracts, decrease
their mass. Although for BPS states the mass will remain constant, the same
physical effect will reduce the string size. The idea is to deform the
ensemble at $g=0$ by adding to the entropy $S_{0}(N,{\mathcal R})$ a term
describing the gravitational interaction between the $\sqrt{N}$ bits of the
polymer, so that 
\[
S_{g}\left( N,{\mathcal R}\right) \sim 4\pi \sqrt{N}-\frac{\pi }{8}\sqrt{N}\,%
\frac{\alpha ^{\prime }}{{\mathcal R}^{2}}+c\,g^{2}N\frac{\sqrt{\alpha
^{\prime }}}{{\mathcal R}}\ ,
\]%
for large $N$ and $\alpha ^{\prime }\ll {\mathcal R}^{2}\ll \alpha ^{\prime }%
\sqrt{N}$. The four--dimensional gravitational character of the  interaction
fixes the ${\mathcal R}$ dependence and the $g^{2}$ factor in the last term.
Finally, $c$ is a positive number of order one. The value of ${\mathcal R}$
that maximizes $S_{g}$ is given by 
\begin{equation}
\langle{\mathcal R}^{2}\rangle\sim \frac{\alpha ^{\prime }}{g^{4}N}\ .  \label{SIsize}
\end{equation}%
This should be a rough estimate of the string size for $N^{-3/4}\ll g^{2}\ll
N^{-1/2}$. 

We conclude that, as we vary the coupling constant $g$, the string has three
different phases. For $g^{2}\ll N^{-3/4}$ the string is well described by a
free polymer and $\langle{\mathcal R}^{2}\rangle\sim \alpha ^{\prime }\sqrt{N}$. In the
intermediate region $N^{-3/4}\ll g^{2}\ll N^{-1/2}$, the string is well
described by a self--interacting polymer and its average size is given by (%
\ref{SIsize}). For $g^{2}\gg N^{-1/2}$ the string becomes a black hole with
horizon radius squared ${\mathcal R}_h^{\,2}=\alpha ^{\prime }/2$, as explained in
Section 2. The three phases are schematically plotted as a function of the
coupling in Figure \ref{fig1} of the Introduction.

\section{Conclusion}

In this work we studied the string/black hole transition in the controlled setting of BPS
string states of the Heterotic string. We have shown evidence that, by increasing the string coupling,
free winding string states evolve into small black holes. There are several computations that would
confirm the physical picture here discussed.

The gravity description, including all $\alpha'$ corrections, is exact in the classical limit $g \rightarrow 0$ with
the classical charges $\rho_n$ and $\rho_m$ fixed. In this limit the  coupling $\lambda=g^2\sqrt{N}$
is finite and $N\rightarrow \infty$. Therefore the free string regime corresponds to $\lambda\equiv 0$ and infinite
string size, as can be seen from Figure \ref{fig1}. On the other hand, the self--gravitating string phase, for which
$\lambda\ll 1$, should be seen in the gravity framework. This fact allows for a direct test of the string/black
hole transition by studying the interpolating solution between the asymptotic flat spacetime (\ref{hole})
and the $AdS_2\times S^2$ near--horizon geometry (\ref{nearhorizon}). For large enough $\lambda$ the black hole solution exists and has a 
\textit{charge independent} horizon radius in the string frame. Below some critical value $\lambda_c \sim 1$  
we enter the self--gravitating string  domain and no longer expect the interpolating solution to exist.
Note that, when $\lambda\sim \lambda_c$, the asymptotic value of the coupling at infinity is of the order
of the fixed value at the horizon.  Moreover the string coupling will always start to decrease as we move towards 
the black hole. Therefore, if an interpolating solution exists for $\lambda\ll\lambda_c$ its string coupling must 
be a non--monotonic function of the radial variable. These seems unlikely to us even though we do not have
a compelling argument. It would be very nice to check this conjecture.

Another check of the transition is to determine the behaviour of the order parameter $\sigma_{abs}(\omega=0)$
away from the free string phase. This would require the computation of the 1--loop absorption amplitude. It would be nice 
to see the splitting of the universal result $G_N S$ for different polarizations, as schematically presented in 
Figure \ref{fig2}.

There are excitations of the BPS string associated to the creation of $NS5/\overline{NS5}$ brane pairs that
are important when considering the near extremal version of the system \cite{Mathur:2005ai}. In fact, the mass
gap for these excitations  decreases as the gravitational coupling is increased. 
Close to (or possibly at) the string/black hole transition the $NS5/\overline{NS5}$ pairs mass gap crosses 
the fundamental string excitations 
mass gap. This could mean that in our computations of the cross section the creation of $NS5/\overline{NS5}$ pairs
should have been considered. However, this is not the case. For finite
energies of the incoming particle, the fundamental string excitations are entropically favoured so that they
dominate the dynamics of the absorption process. On the other hand, when considering the {\em classical} near extremal
version of the black hole the $NS5/\overline{NS5}$ excitations are favoured. In this limit the energy $\Delta E$ 
above the BPS ground state is infinite, with a fixed ratio with respect to the black hole mass. It would be
very interesting to study the near extremal version of this black hole and corresponding transition.

\begin{center} 
{\bf Acknowledgements} 
\end{center}
We wish to thank Costas Bachas for many discussions and for collaboration at the early stages of this work.
Our research is supported in part by INFN, by the MIUR--COFIN contract
2003--023852, by the EU contracts MRTN--CT--2004--503369,
MRTN--CT--2004--512194, by the INTAS contract 03--51--6346, by the NATO
grant PST.CLG.978785 and by the FCT contract POCTI/FNU/38004/2001.
L.C. is supported by the MIUR contract ``Rientro dei cervelli'' part VII.
M.C. is partially supported by the {\em Funda\c{c}\~ao para a Ci\^encia e Tecnologia} (FCT)
grant SFRH/BSAB/530/2005. J.~P. and P.V. are funded by the FCT fellowships 
SFRH/BD/9248/2002 and SFRH/BD/17959/2004/0WA9.
\emph{Centro de F\'{i}sica do Porto} is partially funded by FCT
through the POCTI programme.

\section*{A. Barrier penetration at low energies}

In this appendix we will study the problem of barrier penetration for very small energy.
We will present a general method to compute the small energy behaviour of the transmission amplitude 
through a one--dimensional potential $V$ which tends to zero at infinity.

\begin{figure} 
\centering\includegraphics[width=15cm]{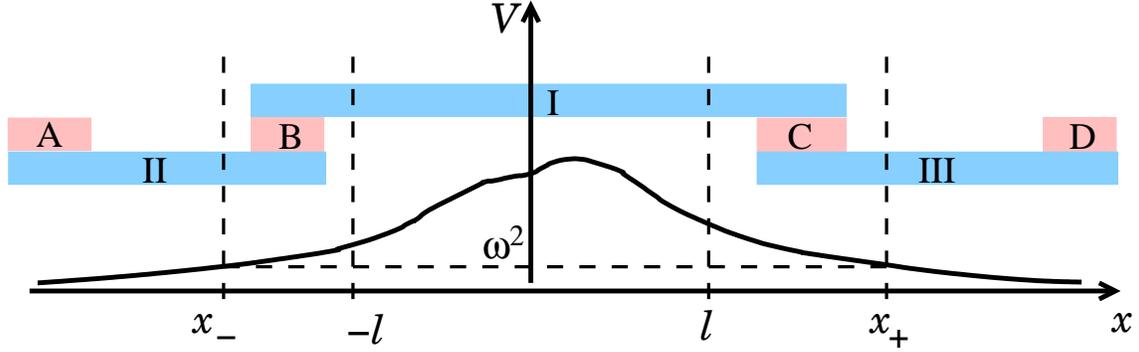} 
\caption{\small{ Schr\"odinger potential. The classical turning points are $x_+$ and $x_-$ and $l$ is the largest length
scale of the potential. Region I is defined by $\omega^2 \ll V(x) $ so that one can set $ \omega=0$ in the  Schr\"odinger equation.
Regions II and III are the regions where the potential is well approximated by its dominant asymptotic behaviour.
Regions B and C are the intersection of region I and regions II and III, respectively. In regions A and D the potential
is much smaller than $\omega^2$.}}
\label{fig4} 
\end{figure}

The Schr\"odinger equation for a wave function $\Psi$ is
$$
\left( \omega^2 + \frac{d^2}{dx^2} \right) \Psi(x) = V(x)\, \Psi(x) \ .
$$
The general idea will be to solve this equation using different approximations in different regions and then
glue the various solutions in the overlapping zones.
Firstly, let us define region I (see Figure \ref{fig4}) by the region where one can neglect the $\omega^2  $ term in 
Schr\"odinger equation.
This can be done as long as $\omega^2 \ll V(x) $ or equivalently for $x_- \ll x \ll x_+ $, where $x_{\pm}$ are the 
classical turning points defined by $V ( x_{\pm}  )=\omega^2 $. Notice that the size of this
region diverges as $\omega\rightarrow0$ because  the potential goes to zero at infinity.
Secondly, if $l$ is the largest length scale of the potential $V$, then for $|x|\gg l $ the potential is well described by
its dominant asymptotic behaviour. We shall call these regions II and III, respectively for negative and positive $x$.
For small enough $\omega$ the classical turning points, $x_{\pm}$, are deep inside regions II and III and one has 
large overlapping zones B and C with region I. We also define regions A and D as the regions where $V(x) \ll \omega^2$ 
(see Figure \ref{fig4}). Let us remark that, the following analysis is general enough to allow for 
\textit{any} linear equation of motion in the middle of region I, not necessarily of the Schr\"odinger type.

For each region A, B, C and D there are two natural independent solutions of the Schr\"odinger equation.
In the asymptotic regions A and D the natural choice are the WKB wave functions
\begin{eqnarray*}
&&\Psi_1^{A,\,D}(x)  \simeq\sqrt{ \frac{\omega}{p(x)}} \,\exp \left( i\int_{x_{\mp}}^x p(x')dx' \right) \ , 
\spa{0.4}\\
&&\Psi_2^{A,\,D} (x)\simeq \sqrt{ \frac{\omega}{p(x)}} \,\exp \left( -i\int_{x_{\mp}}^x p(x')dx' \right) \ ,
\end{eqnarray*}
where $p(x)\equiv\sqrt{\omega^2 -V(x)}$.  
The scattering problem is to find the matrix $M$ such that
$$
\Psi_i^A = M_{ij} \, \Psi_j^D \ .
$$
Then 
$$
M = \left(
\begin{array}{ccc}
\displaystyle{\frac{1}{T} \spa{0.5}} 
& \displaystyle{\frac{R}{T}} 
\\
\displaystyle{   \frac{R^*}{T^*}   } 
& \displaystyle{  \frac{1}{T^*} }   
\end{array}
\right)\ ,
$$
where $T$ and $R$ are the transmission and reflection coefficients which satisfy $ |T|^2+|R|^2=1$.
Our method consists in writing $M$ as the product $M^{II}\, M^{I}\, M^{III} $ where
$$
\Psi_i^A = M^{II}_{ij} \, \Psi_j^B \ , \ \ \ \ \ \ \ \ \
\Psi_i^B = M^{I}_{ij} \, \Psi_j^C \ , \ \ \ \ \ \ \ \ \
\Psi_i^C = M^{III}_{ij} \, \Psi_j^D \ .
$$ 

The natural independent solutions $\Psi^{B,\,C } $ in regions B and C depend on the asymptotic form of the potential
but not on the energy $\omega^2$ because in these regions the energy is negligible compared to the potential.
Therefore the matrix  $M^{I}$ is also independent of $\omega$ since
it is obtained from solving  Schr\"odinger equation for $\omega=0$  and gluing  $\Psi^B$ with $\Psi^C$.
On the other hand,  $M^{II}$ and $M^{III}$ depend on $\omega$ and only on the asymptotic behaviour
of the potential. Thus, we see that the low energy behaviour of the transmission coefficient is
essentially controlled by the asymptotic form of the potential.
We now study the Schr\"odinger equation in regions II and III for different asymptotic behaviours of the potential $V$.

We consider first the case where $V(x)$ decays faster than $1/x^2$ at infinity.
In this case the asymptotic wave functions $\Psi^A$ and $\Psi^D$ are just the usual plane waves
$$
\Psi_1^{A,\,D}(x) \simeq e^{i\omega x} \ , \ \ \ \ \ \ \ \ \ \ 
\Psi_2^{A,\,D} (x) \simeq e^{-i\omega x}\ ,
$$
up to some phase shift that goes to zero as $\omega$ tends to zero.
Studying the asymptotics of the Schr\"odinger equation with $\omega=0$ it is easy to see that the natural independent solutions
 in regions B and C have the following leading behaviour
$$
\Psi_1^{B,\,C}(x) \simeq 1 \ , \ \ \ \ \ \ \ \ \ \ 
\Psi_2^{B,\,C} (x) \simeq x \ .
$$
For these potentials, the distance between $0$ and  the classical turning points $x_{\pm}$ grows slower 
than the asymptotic wave length $2\pi/\omega$ as $\omega$ tends to zero.
The size of the transition interval  $[a,b]$, where the potential influences the wave function between regions A and B
(or regions  C and D), is of the order of  $x_-$ (or $x_+$).
This means that the wave function essentially keeps its asymptotic plane wave form until region B (or C).
Another way to see this is to estimate the variation in the derivative of the wave function through the transition interval
$$
\left. \frac{d\Psi }{dx }\right|_b   - \left. \frac{d\Psi }{dx }\right|_a =
\int_a^b \left(V(x)-\omega^2 \right)\Psi(x)dx \sim \omega^2 x_- \ll \omega \sim \left. \frac{d\Psi }{dx }\right|_a \ .
$$ 
Furthermore, regions B and C from the point of view of the plane waves are the same as $x \simeq 0$, 
so that the matching can be done using $\exp(\mp i\omega x) \simeq 1 \mp i\omega x $, yielding
$$
M^{II} = \left(
\begin{array}{ccc}
\displaystyle{1} 
& \displaystyle{i\omega } 
\\
\displaystyle{   1   } 
& \displaystyle{  -i\omega }   
\end{array}
\right)\ ,
$$
and
$$
M^{III} = \left( M^{II} \right)^{-1}= \frac{1}{2}\left(
\begin{array}{ccc}
\displaystyle{1} 
& \displaystyle{1 } 
\\
\displaystyle{  -i/\omega    } 
& \displaystyle{i/\omega }   
\end{array}
\right)\ .
$$

If the potential decays slower than $1/x^2$ the WKB approximation is applicable \cite{Landau}.
Away from the turning points the WKB approximation requires
$$
|V'(x)| \ll | p(x)|^3 \ ,
$$
where prime stands for $\frac{d}{dx}$.
This condition is always satisfied in regions A and D, however, only if the potential decays slower than $1/x^2$
it is also verified in regions B and C.
In this case, we choose
\begin{eqnarray*}
&&\Psi_1^{B,\,C}(x) \simeq V^{-1/4}(x)\, \exp \left( \pm \int_0^x dx' \sqrt{V(x')} \right) \ , 
\spa{0.4}\\
&&\Psi_2^{B,\,C} (x) \simeq  V^{-1/4}(x)\, \exp \left( \mp \int_0^x dx' \sqrt{V(x')} \right)  \ .
\end{eqnarray*}
To glue the solutions across the turning points $x_{\pm}$ we
use the standard linear approximation of the potential $V(x)\simeq \omega^2 + V'(x_{\pm}) (x-x_{\pm}) $
and solve the Schr\"odinger equation for this linear potential. 
To be able to use the linearized potential to glue the WKB wave functions on the two sides of the turning point,
we need 
$$
\left|V''(x_{\pm}) \right|  \ll  \left| V'(x_{\pm}) \right|^{4/3} \ .
$$  
In the small $\omega$ limit this is equivalent to the fact that the potential decays slower than $1/x^2$
as $x$ goes to infinity\footnote{Here we are assuming that the potential is not pathological like for example $x^{-1}\sin^2(x^2)$.}.
The general solution around the turning point $x_- $ is given by a linear combination of the Airy functions
$Ai(z)$ and $Bi(z)$, where $z=|V'(x_-)|^{1/3}(x-x_-)$.
Matching the asymptotics of the Airy functions with the WKB wave functions close to the turning point we obtain
$$
\Psi_{1,2}^A(x)  \simeq \sqrt{ \pi \omega } \,|V'(x_-)|^{-1/6} \, e^{ \pm i \pi /4}
\left( Bi(z) \mp i Ai(z) \right) \ ,\ \ \ \ \ \ \ \ \ \ 
x \simeq x_- \ ,
$$
and
\begin{eqnarray*}
&&\Psi_1^B(x) \simeq e^{-P_-}  \,|V'(x_-)|^{-1/6}  \,\sqrt{\pi} Bi(z)        \ , 
\spa{0.3}\\
&&\Psi_2^B (x) \simeq  e^{P_-}  \,|V'(x_-)|^{-1/6}  \, 2\sqrt{\pi} Ai(z)   \ ,
\ \ \ \ \ \ \ \ \ \ x  \simeq x_- \ ,
\end{eqnarray*}
where $P_-=\int_{x_-}^0 dx' |p(x')|  $.
Thus,
$$
M^{II} =  e^{-i\pi/4} \,\frac{\sqrt{\omega}}{2}\left(
\begin{array}{ccc}
\displaystyle{2 i  e^{P_-}         } 
& \displaystyle{      e^{-P_-}          } 
\\
\\
\displaystyle{   2e^{P_-}        } 
& \displaystyle{  i   e^{-P_-}   }   
\end{array}
\right)\ .
$$
Similarly
$$
M^{III} = \frac{e^{-i\pi/4}}{2\sqrt{\omega}}\left(
\begin{array}{ccc}
\displaystyle{   i  e^{-P_+}         } 
& \displaystyle{        e^{-P_+}            } 
\\
\\
\displaystyle{   2e^{P_+}      } 
& \displaystyle{ 2 i e^{P_+}   }   
\end{array}
\right)\ ,
$$
with $P_+ = \int_0^{x_+} dx' |p(x')|  $.

Finally, we consider the marginal situation where the potential has the asymptotic form
$$
V(x) \sim \frac{U_{\pm}}{x^2} \ ,\ \ \ \  \ \ \ \ \ \ \ x \rightarrow \pm \infty \ .
$$
In this case, $\Psi^A$ and $\Psi^D$ tend to plane waves with an extra phase
$$
\Psi_1^{A,\,D}(x) \simeq \exp \left( i\omega x \pm i\, \frac{\pi}{2}\, \sqrt{U_{\mp}}  \right) \ , \ \ \ \ \ \ \ \ \ \ 
\Psi_2^{A,\,D} (x) \simeq  \exp \left(  -i\omega x  \mp i\, \frac{\pi}{2}\, \sqrt{U_{\mp}} \right)   \ .
$$
The natural choices for $\Psi^B$ and $\Psi^C$ are
$$
\Psi_1^{B,\,C}(x) \simeq x^{1/2 - \beta_{\mp}}  \ , \ \ \ \ \ \ \ \ \ \ 
\Psi_2^{B,\,C} (x) \simeq   x^{ 1/2 + \beta_{\mp} }    \ ,
$$
where $U_{\pm}=\beta_{\pm}^{\,2} -1/4  $.
For this specific asymptotic behaviour of the potential the solutions in regions II and III can be written 
in terms of Bessel functions. Using the asymptotic expansions  of the Bessel functions, one determines the wave
functions in region II
\begin{eqnarray*}
&&\Psi_{1,2}^A(x) \simeq  - \frac{e^{\,\pm i\phi_-^-} }{\sin(\beta_- \pi)}\,
 \sqrt{ \frac{\pi\omega x}{2}  } \,\left[ J_{\beta_-}(\omega x) - e^{\pm i \beta_- \pi} J_{-\beta_-}(\omega x)\right] \ , 
\spa{0.4}\\
&&\Psi_{1,2}^B (x) \simeq \Gamma(1 \mp \beta_-) \left( \frac{2}{\omega}\right)^{1/2 \mp \beta_-}
 \sqrt{ \frac{\omega x}{2}  } \,J_{\mp \beta_-}(\omega x) \ ,
\end{eqnarray*}
where we define $\phi_a^\pm= \frac{\pi}{2}( \sqrt{U_a } \pm \,\beta_a - 1/2)  $ with the index $a=\pm$.
This yields
$$
M^{II} =\left(
\begin{array}{ccc}
\displaystyle{ e^{ i\phi_-^+} f(\beta_-)  } 
& \displaystyle{     e^{i\phi_-^- } f(-\beta_-)  } 
\\
\\
\displaystyle{   e^{ -i\phi_-^+} f(\beta_-) } 
& \displaystyle{  e^{ - i\phi_-^-}f(-\beta_-) }   
\end{array}
\right) \ ,
$$
with $f(\beta)= \frac{1}{\sqrt{\pi}} \Gamma(\beta)   \left( \frac{\omega}{2}\right)^{\frac{1}{2} - \beta} $.
A similar calculation gives
$$
M^{III} =\frac{1}{2}  \left(
\begin{array}{ccc}
\displaystyle{  e^{ i\phi_+^+} f(1-\beta_+) } 
& \displaystyle{    e^{ -i\phi_+^+} f(1-\beta_+)} 
\\
\\
\displaystyle{  e^{ i\phi_+^-}  f(1+\beta_+) } 
& \displaystyle{ e^{ -i\phi_+^-} f(1+\beta_+) }   
\end{array}
\right)\ .
$$

In all cases, the dominant low energy behaviour of the transmission probability is given by
\beq
|T|^2 \simeq \left| M^{II}_{11} \right|^{-2} \left| M^{III}_{21} \right|^{-2} \left| M^{I}_{12} \right|^{-2} \ ,
\label{transmission}
\eeq
from which we conclude that
$$
|T|^2 \sim \omega^{2\beta_- +2\beta_+} \ ,
$$
if the potential decays as $1/x^2$. 
If it decays faster than $1/x^2$ for $x \rightarrow \pm \infty$, one should replace $\omega^{2\beta_{\pm}}$ by
$\omega$. If it decays slower than $1/x^2$, the factor $\omega^{2\beta_{\pm}} $ should be replaced by 
$\exp\left( \mp  2\int_0^{x_{\pm}}dx'|p(x')| \right)$. 
If it decays slower on both sides one recovers the usual WKB tunneling factor
$$
|T|^2 \sim \exp \left( -2   \int_{x_-}^{x_+}dx'\sqrt{V(x') - \omega^2  }        \right) \ .
$$
As we have just seen, the $\omega$ dependence of the low energy transmission probability is independent of the details of the potential,
uniquely relying on its asymptotics. 
However, if one is interested in the precise numerical factors, it is enough to 
determine $\left| M^{I}_{12} \right|^{2} $ for some given potential and use
formula (\ref{transmission}).
It is then clear that this method allows \textit{any} linear evolution
in the middle region since we only need the $\omega$ independent numerical coefficient $M^{I}_{12}$.

\section*{B. Averages over string states}

In the main text we needed the average occupation number $\langle r^l_n \rangle $, in the ensemble of all partitions
of the integer $N$ in integers of $24$ colours. More precisely, an element of this ensemble 
is the set of non--negative integers $\{r^i_m \} $ obeying the constraint
\beq
\sum_{i=1}^{24} \sum_{m=1}^{\infty}  m\, r^i_m = N \ . 
\label{constraint}
\eeq
We are interested in the average $\langle  r^l_n \rangle_N$ for fixed and large $N$. In this limit,
 one can trade the constraint (\ref{constraint}) by a chemical potential $\mu$
$$
\langle  r^l_n \rangle_{\mu}=\frac{1}{Z} \sum_{ \{r^i_m \} } r^l_n e^{-\mu \sum_{i=1}^{24} \sum_{m=1}^{\infty}  m\, r^i_m }
= \frac {1}{e^{\mu n}-1} \ ,
$$
where $Z= \sum_{ \{r^i_m \} } \exp\left(-\mu \sum_{i=1}^{24} \sum_{m=1}^{\infty}  m\, r^i_m \right)$ is the usual partition function.
In the limit of large $N$ (or small $\mu$) the averages $\langle  \  \rangle_{\mu} $ and $\langle  \  \rangle_{N} $ are related
through
$$
\langle  N \rangle_{\mu}= 24 \sum_{n=1}^{\infty}\frac {n}{e^{\mu n}-1} \simeq 
\frac{24}{\mu^2} \int_0^{\infty}\frac {x\,dx}{e^{x}-1}=  \frac{4\pi^2}{\mu^2} \ .
$$
Thus
$$
\langle  r^l_n \rangle_N= \frac {1}{\exp\left(\frac{2\pi n}{\sqrt{N}}\right)-1} \ .
$$

Following \cite{Damour} we will estimate the number of string states with a given size
$$
{\cal R}^2=\alpha' \sum_{m=1}^{\infty}  \frac{1}{m}\, r^i_m \ .
$$ 
To this end we introduce a chemical potential  $\gamma$ conjugated to the size of the string.
Defining 
$$
e^{F(\mu,\gamma)} \equiv  \sum_{ \{r^i_m \} } e^{-\mu N -\gamma {\cal R}^2/\alpha'} \ , 
$$
we obtain the entropy as a function of ${\cal R}$ from the Legendre transform
$$
S(N,{\cal R})=F(\mu,\gamma)+\mu N +\gamma {\cal R}^2/\alpha' \ ,
$$
where $N$ and ${\cal R}$ are related to $\mu$ and $\gamma$ through
$$
N=-\frac{\partial F }{\partial \mu } \ ,\ \ \ \ \ \ \ \ \ \ \ \ \ \ \
{\cal R}^2=-\alpha'\frac{\partial F }{\partial \gamma } \ .
$$
The free energy $F$ is then given by
\beq
F(\mu,\gamma)=-23 \sum_{n=1}^{\infty}\log\left(1-e^{-\mu n}\right) - \sum_{n=1}^{\infty}\log\left(1-e^{-\mu n -\gamma/n}\right) \ .
\label{F}
\eeq
Let us first consider the region of parameters $\mu \ll \sqrt{\mu\gamma} \ll 1$.
In this case the sums in $F$ can be well approximated by integrals
$$
F(\mu,\gamma)\simeq -\frac{23}{\mu} \int_0^{\infty} dx \log\left(1-e^{-x}\right) 
- \frac{1}{\mu} \int_0^{\infty} dx\log\left(1-e^{-x -\mu\gamma/x}\right) \ ,
$$
so that
$$
F(\mu,\gamma)\simeq \frac{4\pi^2}{\mu} -\pi \sqrt{\frac{\gamma}{\mu}}  + O(\gamma) \ .
$$
The entropy is then
$$
S(N,{\cal R})= 4 \pi \sqrt{N} \left( 1 - \frac{\alpha'}{32{\cal R}^2} \right) \ ,
$$
for large $N$ and $\alpha' \ll {\cal R}^2 \ll \alpha'  \sqrt{N}$.

To determine the entropy function for values of ${\cal R}^2$ larger than the typical value $\alpha' \sqrt{N}$
we need to consider negative chemical potential $\gamma$. The free energy (\ref{F}) is well defined
as long as $\gamma >-\mu$. For $0<\gamma+\mu \ll \mu \ll 1$\ ,
$$
F(\mu,\gamma)\simeq \frac{4\pi^2}{\mu} -\log( \gamma+ \mu)  + O(1) \ ,
$$
where the $\gamma$ dependence steams from the first term in the sum. 
Therefore
$$
S(N,{\cal R})= 4 \pi \sqrt{N} \left( 1 - \frac{{\cal R}^2}{2\alpha'N} \right) \ ,
$$
for large $N$ and $\alpha'\sqrt{N} \ll {\cal R}^2 \ll \alpha'  N$.

\section*{C. Average string size}

In this appendix we compute the average size of the string in the spatial direction $X^{\mu}$  $(\mu=1)$
of the null vector ${\bf k}$ appearing in the DDF operators. In this case we have different commutation
relations
$$
\left[\alpha^{\mu}_{m}, A^i _n \right]=n \sqrt{\frac{\alpha'}{2}}\, k^{\mu} D_{m,n}^i  \ , \ \ \ \ \ \ \ \ \ \ \ \ \ \ 
\left[ D_{m,n}^i, A^j_l \right]=-l\, \delta^{ij} \, E^{n+l}_m   \ ,
$$
where
$$
D_{m,n}^i =\sqrt{\frac{2}{\alpha'} } \oint  \frac{dz}{2\pi } \, z^{m} \, \partial X^i(z) \,e^{\,i\,n\,{\bf k} \cdot {\bf X}_{L}(z) }\ ,
$$ 
and
$$
E^n_m =\oint  \frac{dz}{2\pi } \, z^{m} \,{\bf k} \cdot \partial {\bf X}_{L}(z) \,e^{\,i\,n\,{\bf k} \cdot {\bf X}_{L}(z) }\ .
$$ 
Furthermore
$$
\left[D_{-m,n}^i ,D_{m,-l}^j \right]= \delta^{ij} \left( n \delta_{n,l} -m B_0^{n-l} \right)  \ , \ \ \ \ \ \ \ \ \ \ \ \ \ \ 
\left[E^n_m, A^i_l \right]=  \left[E^n_m,D_{m,l}^i \right] =0   \ .
$$
Commuting the $\alpha^{\mu}_{m} $ we obtain
$$
\Delta_m =\frac{\alpha'}{2}\left( k^{\mu}\right)^2  \left|\sum_i \sum_{n=1}^{\infty}   \sum_{s=1}^{r^i_n}
  \prod_{ l=1}^{n-1}  \frac{  \left(   A^{i} _{-l}   \right)^{r^{i}_l}  }
{\sqrt{ r^{i}_l! \, l^{ r^{i}_l}    } } \,
 \frac{  \left(   A^{i} _{-n}   \right)^{r^{i}_n -s}   n  D_{m,-n}^i      \left(   A^{i} _{-n}   \right)^s  }
{\sqrt{ r^{i}_n! \, n^{ r^{i}_n}    } }
 \!\prod_{ l=n+1}^{\infty} \!\frac{  \left(   A^{i} _{-l}   \right)^{r^{i}_l}  }
{\sqrt{ r^{i}_l! \, l^{ r^{i}_l}    } }
   | {\bf p}_L + N {\bf k}  \rangle 
\right|^2 \!.
$$
Since $   D_{m,-n}^i    | {\bf p}_L + N {\bf k}  \rangle =0$ for $n<m$ we commute the $ D_{m,-n}^i  $ with $n<m$
to the right and the ones with  $n>m$ to the left.
In this process we will generate many  $E_m^{-n-l}$ operators. However, because these commute with all the other operators 
and satisfy $ E_m^{-n} | {\bf p}_L + N {\bf k}  \rangle =0 $ for  $n<m$  and 
$ \left( E^{-n} _m \right)^{\dagger}   =   E^{n} _{-m} $, only the $E^{-m}_m$ operators remain.
The contribution of the latter is easily computed using
 $ E^{-m} _m  | {\bf p}_L + N {\bf k}  \rangle = | {\bf p}_L + (N-m) {\bf k}  \rangle$.
Finally, using $ D_{m,-m}^i  | {\bf p}_L + N {\bf k}  \rangle = \sqrt{\alpha'/2}\,  p_L^i\, | {\bf p}_L + (N-m) {\bf k}  \rangle$,
we arrive at
$$
\barr{c}
\displaystyle{\Delta_m =
\frac{\alpha'}{2}\left( k^{\mu}\right)^2\,\sum_i \left( \frac{\alpha'}{2}\left( p_L^i \right)^2 m r^i_m +
\!\!\sum_{n+l=m} n l r^i_n r^i_l + \!\!\sum_{n=m+1}^\infty n  r^i_n (n-m) + \!\!\sum_{n-l=m} n l r^i_n r^i_l 
 \right)=}\spa{0.7}\\
\displaystyle{mr^4_m + 
\frac{2}{\alpha'M^2}  \left[  \,\sum_i \left( 
\sum_{n+l=m} n l r^i_n r^i_l + \!\!\sum_{n=m+1}^\infty n  r^i_n (n-m) + \!\!\sum_{n-l=m} n l r^i_n r^i_l 
 \right) 
 -2(N-1) mr^4_m \right]\ .}
\earr
$$
After averaging $ \Delta_m$ over the microstates one should obtain the same result as for the other directions,
$$
\langle \Delta_m \rangle =m  \langle r^i_m \rangle   \ ,
$$
for any $N$ and $m$. This implies the identity
\beq
\sum_{n=1}^m n (m-n) \langle r^i_n r^i_{m-n}\rangle + \sum_{n=m+1}^\infty n (n-m)  \langle r^i_n  ( r^i_{n-m}    +1) \rangle 
= \frac{N-1}{12} \,m\langle r^i_m\rangle \ .
\label{identity}
\eeq
In the limit $m\ll \sqrt{N}$,  it is easy to see that both sides of (\ref{identity}) have the same
dominant behaviour $   N^{3/2}/( 24 \pi ) $.
This ensures that  
$$
\langle \Delta_m \rangle \simeq \frac{\sqrt{N}}{2\pi}   \ ,
$$
as expected.

\end{document}